\documentclass[12pt]{iopart}
\usepackage{iopams}

\usepackage{array}
\usepackage{graphicx} 
\usepackage{color}
\usepackage{cite}
\usepackage{epstopdf}

\newcommand{\bra}[1]{\langle #1 |}
\newcommand{\ket}[1]{| #1\rangle}
\newcommand{\braket}[2]{\langle #1 | #2\rangle}

\newcolumntype{L}{>{$}l<{$}}
\begin{document}

\title{R\'{e}nyi entropy of the totally asymmetric exclusion process}

\author{Anthony J. Wood, Richard A. Blythe, Martin R.  Evans}

\address{School of Physics and Astronomy, University of Edinburgh, Peter Guthrie Tait Road, Edinburgh EH9 3FD}

\begin{abstract}
 The R\'{e}nyi entropy is a generalisation of the Shannon entropy that is sensitive to the fine details of a probability distribution. We present results for the R\'{e}nyi entropy of the totally asymmetric exclusion process (TASEP). We calculate explicitly an entropy whereby the \emph{squares} of configuration probabilities are summed, using the matrix product formalism to  map the problem to one involving a six direction lattice walk in the upper quarter plane. We derive the generating function across the whole phase diagram, using an \emph{obstinate kernel} method. This gives the leading behaviour of the R\'{e}nyi entropy and corrections in all phases of the TASEP. The leading behaviour is given by the result for a Bernoulli measure and we conjecture that this holds for all R\'{e}nyi entropies. Within the maximal current phase the correction to the leading behaviour  is logarithmic in the system size. Finally, we  remark upon a special property of equilibrium systems whereby discontinuities in the R\'{e}nyi entropy arise away from phase transitions, which we refer to as secondary transitions. We find no such secondary transition for this nonequilibrium system, supporting the notion that these are specific to equilibrium cases.
\end{abstract}

\maketitle

\section{Introduction}
\label{S:intro}

The essence of statistical mechanics is that the macroscopic properties of a many-body system are determined by the probability distribution over its microstates. Given the distribution $p_i$, where $i$ labels the microstates, one can construct the familiar Shannon entropy
\begin{equation} \label{eq:Shannon}
S= -\sum_{i} p_i \log{p_i}
\end{equation}
that specifies the level of microscopic uncertainty implied by the system's macrostate.

The \emph{R\'{e}nyi entropy} \cite{Renyi1961} is less well known. This is a one-parameter deformation of \eref{eq:Shannon} that is sensitive to the details of the probability distribution and is defined as
\begin{equation}\label{eq:Renyi}
H_\lambda = \frac{1}{1-\lambda} \log\sum_{i} p_i^\lambda \;.
\end{equation}
Although valid for any base of logarithm, we will use the natural logarithm throughout this work. This entropy is structurally very similar to the Shannon entropy, which is recovered as $\lambda \to 1$ (see \eref{eq:limto1}). It varies continuously with the probabilities, is invariant under relabelling of the microstates, and is maximised by a probability distribution that is uniform over the entire state space. Most importantly, upon combining independent systems, the R\'{e}nyi entropy is additive  \cite{Renyi1961}. As we will discuss below (see also \cite{Baez2011}), the entire family of R\'{e}nyi entropies for an equilibrium system with a Boltzmann distribution can be expressed in terms of the equilibrium free energy at different temperatures. 

In this work, we are interested in the properties of the R\'{e}nyi entropies in the context of a \emph{nonequilibrium} steady state (NESS). Specifically, we aim to determine how such features as long-range correlations \cite{Schmittmann1995,Hinrichsen2000} and nonequilibrium phase transitions \cite{Evans2005,Henkel2008} manifest themselves in statistical measures like the R\'{e}nyi entropy. We are particularly interested in cases where these measures behave differently in equilibrium and nonequilibrium states.

To this end, we study the R\'{e}nyi entropy for a paradigmatic example of a NESS for which the steady state distribution can be computed exactly, the totally asymmetric exclusion process (TASEP) with open boundaries \cite{Macdonald1968,Derrida1992,Derrida1993a,Schutz1993,Blythe2007,Chou2011}. This system comprises interacting particles moving stochastically across a one-dimensional lattice and nonequilibrium phase transitions are induced by changing the two boundary parameters of the system. For general values of these parameters, the steady state of the TASEP can be represented through a matrix product formalism \cite{Derrida1993a}. From this, an exact phase diagram is derived by analysis of the nonequilibrium partition function (see e.g.~\cite{Blythe2007} for a discussion of different methods). Phase transitions in this system are characterised by  changes in the macroscopic forms of  density profile and particle current, both of which are exactly calculable by the matrix product formalism. The matrix product formalism has also allowed quantities such as the moments of the current \cite{Derrida1995,Lazarescu2011,lazarescu2015physicist} to be computed and has been extended to solve open systems with many species of particle \cite{Arita2006,Ayyer2009,Crampe2015,Cantini2016,Crampe2016a,Crampe2016b}.

Calculation of the R\'{e}nyi entropy is technically challenging as it involves raising each of the microstate probabilities $p_i$ to some power $\lambda$. We present an exact calculation for the case $\lambda=2$, which can be achieved using a matrix product representation to  map to a two-dimensional random walk problem with absorbing boundaries. The solution of this problem entails a generalisation of what is known in the mathematical literature as the \emph{obstinate kernel} method for computing generating functions \cite{Bousquet2005,Bousquet2010}. For all three phases of the TASEP, we obtain  the $\lambda=2$ entropy  which provides a phase-space localisation measure, that is, an effective number of participating microstates in each phase.  We find that the leading order behaviour of this entropy corresponds with that of a Bernoulli measure, and that the form of the leading correction reflects the range of the correlations present in each of the phases. Finally, we show that the nonequilibrium phase transitions in the TASEP give rise to an analytical structure of the R\'{e}nyi entropy that distinguishes itself from that seen in equilibrium systems.

\subsection{R\'{e}nyi entropy}\label{SS:Renyi}

For orientation, we discuss in more detail some general properties of the R\'{e}nyi entropy, \eref{eq:Renyi}. Consider a system with configurations $\{i\}$ and associated probabilities $\{p_i\}$, normalised so that $\sum_{i}p_i = 1$.  The Shannon entropy is obtained from \eref{eq:Renyi} in the limit $\lambda \to 1$ as follows:
\begin{equation}
\label{eq:limto1}
\lim_{\lambda \to 1}H_\lambda = 
\lim_{\lambda \to 1} \frac{\log{\sum_{i} p_i {\rm e}^{(\lambda-1) \log p_i}}}{1-\lambda} = -\sum_{i} p_i\log p_i = S \;.
\end{equation}
The R\'{e}nyi entropy is a nonincreasing function of $\lambda$: for $\lambda_1 > \lambda_2$, $H_{\lambda_1} \leq H_{\lambda_2}$. Knowledge of $H_0$ and any R\'{e}nyi entropy $H_{\lambda>1}$ then gives upper and lower bounds on the Shannon entropy \cite{Xu2010}.

By increasing $\lambda$, $H_\lambda$ places more weight on more probable configurations, which is made clear with two extreme cases. $H_0$ is simply a measure (specifically the logarithm) of the number of configurations with $p_i>0$, and $H_\infty$ is a measure of only the largest probability in the set $\{p_i\}$ (or probabilities if there is no single most probable configuration) \cite{Csiszar2008}. Thus, by knowing $H_\lambda$ for different values of $\lambda$, the R\'{e}nyi entropy probes  finer details of a probability distribution than the Shannon entropy alone.

Taking the exponential of the R\'{e}nyi entropy gives
\begin{equation}\label{eq:effnum}
\e^{H_\lambda} = \left[\sum_{i}p_i^\lambda\right]^\frac{1}{1-\lambda} \;.
\end{equation}
To interpret this, consider two extremes of the distribution $\{p_i\}$. For a system whereby a single configuration has probability $1$, $\e^{H_\lambda} = 1$. Conversely, for a system with $M$ equally likely configurations, $\e^{H_\lambda} = M$. Thus we interpret $\e^{H_\lambda}$ as an \emph{effective number} of configurations---or, equivalently, a measure of how localised the system is within its configuration space. In ecology, these effective numbers \eref{eq:effnum} are known as \emph{Hill numbers} \cite{Hill1973} and give measures of the diversity of a biological community \cite{Leinster2012,Chao2012}.

The main section of this paper addresses the $\lambda = 2$ case of the entropy, which we make explicit here:
\begin{equation} \label{eq:H2}
H_2 = -\log\sum_{i} p_i^2 \;.
\end{equation}
This is often referred to in the literature as the \emph{collision entropy} \cite{Wehner2010}, and the corresponding effective number
\begin{equation} \label{eq:eH2}
\e^{H_2} = \frac{1}{\sum_{i}p_i^2}
\end{equation}
as the \emph{inverse participation ratio} \cite{Bera2015,Ishizaki2010} (or in the  context of diversity of a biological system \emph{Simpson's reciprocal index} \cite{Hill1973}). This is a commonly used measure of quantum localisation of a wavefunction $\psi$, where $p_i^2 = |\psi|^4$ \cite{Evers2000,Georgeot1997}. Finally, we mention disordered systems where the free energy landscape breaks into pure states with fluctuating weights \cite{Mezard1987}. The sum of these weights squared, which is itself a random variable, provides  information about the free energy landscape and its distribution is related to the Parisi order parameter function \cite{Derrida1987,Derrida1997}.

As noted above, it is possible to connect the R\'{e}nyi entropy to the thermodynamic properties of an equilibrium system, for which $p_i={\rm e}^{-E_i/{k_BT}}/Z(T)$, where $E_i$ is the energy of microstate $i$, $T$ is the temperature and $Z(T)$ is the partition function. The R\'{e}nyi entropy is readily calculable \cite{Baez2011} as
\begin{eqnarray} 
\label{eq:renyieq}
H_\lambda = \frac{1}{1-\lambda}\log\frac{1}{Z(T)^\lambda}\sum_{i}\e^{-\frac{\lambda E_i}{k_BT}}
 = \frac{1}{1-\lambda}\log{\frac{Z(\frac{T}{\lambda})}{Z(T)^\lambda}} \;.
\label{eq:renyieqZ}
\end{eqnarray}
Intriguingly, $H_\lambda$ involves the ratio of two partition functions $Z$ at \emph{different} temperatures $T$ and $T/\lambda$.  Equivalently, using the definition of the free energy $F= -k_BT \log{Z(T)}$,
$H_\lambda$  is proportional to the free energy difference between the two temperatures, and we can interpret this as the amount of work one can extract from the system between these two temperatures \cite{Baez2011}.

The form \eref{eq:Renyi} of the entropy has consequences for systems that exhibit phase transitions. Suppose there is a transition at $T=T^*$. Then, in the thermodynamic limit, there is a nonanalyticity in the partition function $Z(T^*)$. Consequently, in \eref{eq:renyieqZ} we will find not just the usual nonanalyticity at the critical temperature $T = T^*$, but also a secondary transition at $T = \lambda T^*$, away from the critical temperature. These particular properties of $H_\lambda$ rely on the fact that the temperature $T$ appears in a specific way in the statistical weights. With any deviation away from such a distribution---as occurs in a nonequilibrium steady state---the result \eref{eq:renyieqZ} may no longer apply. As such we consider the equilibrium R\'{e}nyi entropy as a special case.

\subsection{The totally asymmetric exclusion process}
\label{SS:TASEP}

\begin{figure}[t]
\centering
\includegraphics{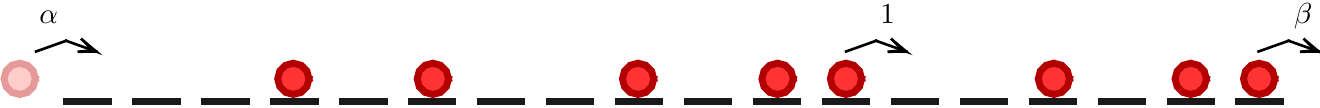}
\caption{The TASEP. Particles enter at rate $\alpha$, hop right at rate $1$, and exit at rate $\beta$. The exclusion property means that particles can not overtake each other.}
\label{F:TASEP}
\end{figure}

We now specify the dynamics and important properties of the totally asymmetric exclusion process (TASEP, Figure~\ref{F:TASEP}) which underpins the rest of this work. The TASEP is a stochastic one-dimensional system, defined on a lattice of $N$ sites. Particles are introduced from a left reservoir at rate $\alpha$ when the first site of the lattice is available, make single hops to the right at unit rate, and are then absorbed from the last site of the lattice by a right reservoir at rate $\beta$. The exclusion property forbids particles from overlapping or overtaking. This traffic-like system approaches a NESS in the long time limit, whereby the density profile and particle current stabilise. 

\begin{figure}[t]
\centering
\includegraphics{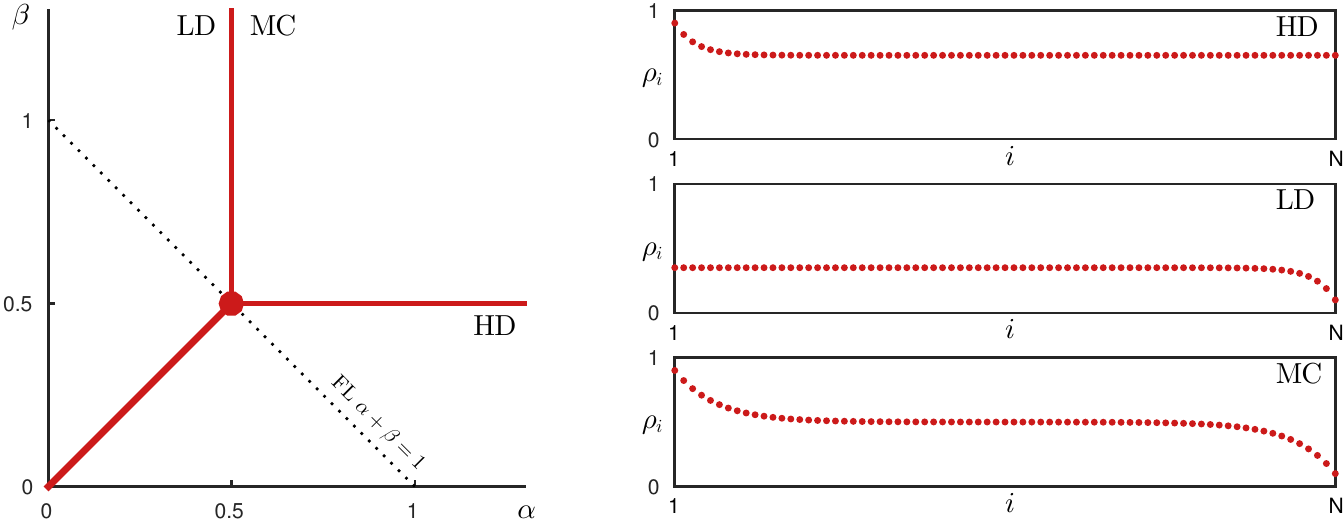}
\caption{Phase diagram of the TASEP. Three main phases - high density (HD), low density (LD), maximal current (MC) - exist, with corresponding typical steady state density profiles $\rho_i$ shown, where $i$ labels the site. The $\alpha + \beta = 1$ factorisation line (FL) is highlighted.}
\label{F:PhaseDia}
\end{figure}

Phase transitions are induced by varying the two reservoir parameters $\alpha$, $\beta$, whereby the bulk density  and current change in a nonanalytic way at the phase boundaries. In physical terms, a different factor limits the particle flow in different parts of the phase diagram (see Figure~\ref{F:PhaseDia}):
\begin{itemize}
\item \emph{Low density (LD)}: $\alpha < \frac{1}{2}$, $\alpha < \beta$. Current is restricted at the left reservoir, and few particles enter the system. The steady state is characterised by a bulk density $\alpha$, and current $J \sim \alpha(1-\alpha)$ .
\item \emph{High density (HD)}: $\beta < \frac{1}{2}$, $\alpha > \beta$. Particles freely enter the system but queue to leave at the right reservoir. The steady state is characterised by a bulk density $1-\beta$, and current $J \sim \beta(1-\beta)$.
\item \emph{Maximal current (MC)}: $\alpha > \frac{1}{2}$, $\beta > \frac{1}{2}$. Particles freely enter and leave the system, so the exclusion interaction in the bulk is what restricts the current. In this phase the bulk density approaches $1/2$ and the current is maximised at $J \sim 1/4$.
\end{itemize}

In the steady state, microstate probabilities in this system are exactly computed by a matrix product formalism (for a comprehensive review, see \cite{Blythe2007}). Briefly, the probability of a configuration of particles ${\cal C}$ is given by an ordered product of matrices, one per site, representative of the sequence of occupied and vacant sites. This product of matrices is reduced to a scalar weight by two vectors, whose relationships with the matrices encode information about the reservoirs.

More precisely, one associates with each particle a matrix $D$ and each vacant site a matrix $E$. The left and right reservoirs correspond to vectors $\bra{W}$ and $\ket{V}$, respectively, with a normalisation chosen such that $\braket{W}{V}= 1$. Then, for example, the weight of a configuration of particles and holes $\mathcal{W}({\cal C} = \_\circ\circ\dots\circ\_)$ is exactly the scalar product $\bra{W}EDD \dots DE\ket{V}$, and the matrix expressions can be reduced to a scalar by repeated use of the reduction rules
\begin{eqnarray}
\label{eq:Reduction}
D\ket{V} = \frac{1}{\beta}\ket{V}\;, & \qquad \bra{W}E = \frac{1}{\alpha}\bra{W}\; , & \qquad DE = D+E \;.
\end{eqnarray}
To scale the weight $\mathcal{W}({\cal C})$ to a probability $P({\cal C})$, there is a normalisation factor
\begin{eqnarray}\label{eq:ZN}
Z_N(\alpha,\beta) &= \sum_{{\cal C}\in\mathcal{L}_N}\mathcal{W}({\cal C}) = \bra{W}(D+E)^N\ket{V}
\end{eqnarray}
where $\mathcal{L}_N$ is the set of $2^N$ TASEP configurations of length $N$. To demonstrate these reduction relations, we calculate the normalisation for $N = 0,1,2$:
\begin{eqnarray}
\fl \bra{W}(D+E)^0\ket{V} = 1 \\
\fl \bra{W}(D+E)\ket{V} = \frac{1}{\beta}+\frac{1}{\alpha} \\
\fl \bra{W}(D+E)^2\ket{V} = \bra{W}(DD+EE+ED+DE)\ket{V} \nonumber \\
\fl = \bra{W}(DD+EE+ED+D+E)\ket{V} = \frac{1}{\beta^2}+\frac{1}{\alpha^2}+\frac{1}{\alpha\beta}+\frac{1}{\beta}+\frac{1}{\alpha}.
\end{eqnarray}
It may be shown by a variety of approaches (see \cite{Blythe2007}, and also Section~\ref{SS:normGF} below), that a closed form expression for $Z_N$ is given by
\cite{Derrida1993a}
\begin{eqnarray}
Z_N &= \sum^N_{p=1}\frac{p(2N-p-1)!}{N!(N-p)!}\left[\frac{\alpha^{-1-p}-\beta^{-1-p}}{\frac{1}{\alpha}-\frac{1}{\beta}}\right] \;.
\label{ZNexact}
\end{eqnarray}

With \eref{eq:Reduction} and the known closed form of the normalisation factor \eref{ZNexact}, configuration probabilities are exactly calculable. However this process of reduction is iterative, in the sense that each ordered product of $D$ and $E$ matrices must be individually evaluated using \eref{eq:Reduction}. Therefore a general expression for $H_\lambda$, analogous to the equilibrium expression \eref{eq:renyieqZ} is difficult to obtain.  With this in mind, we view the $\lambda = 2$ case of $H_\lambda$ the simplest nontrivial entropy to calculate.

There is one line in the phase diagram, $\alpha + \beta = 1$, which we have highlighted in Figure~\ref{F:PhaseDia}, where we can compute $H_\lambda$ straightforwardly. On this line, the reduction relations \eref{eq:Reduction} reduce to the case of $D$ and $E$ commuting, and we may take a scalar representation $D= 1/\beta$ and $E=1/\alpha$ \cite{Derrida1993a, Blythe2007}. We call this the \emph{factorisation line}. Here, configuration probabilities follow a Bernoulli distribution \cite{Blythe2007}, whereby each of the $N$ sites are independently, individually occupied with probability $\rho$. With this, the sum of configuration weights to an arbitrary power $\lambda$ is
\begin{eqnarray}
\sum_{C\in\mathcal{L}_N}P({\cal C})^\lambda &=  (\rho^\lambda+(1-\rho)^\lambda)^N
\end{eqnarray}
and we find the full R\'{e}nyi entropy exactly
\begin{equation}
\label{eq:MFrenyi}
H_\lambda = \frac{N}{1-\lambda}\log{\left(\rho^\lambda+(1-\rho)^\lambda\right)} \;.
\end{equation}
This line traverses the high density phase (where $\rho = 1-\beta$) and the low density phase (where $\rho = \alpha$). Along this line  the statistics of the TASEP here are mean field in nature, with no correlations between neighbouring sites. We note that at the tricritical point $\alpha= \beta = \rho=  1/2$, $H_\lambda  =N\ln 2$ independent of $\lambda$.

\subsection{Sum of squared weights and generating function}

We now move onto the calculation of $H_2$ \eref{eq:H2} for general $\alpha$ and $\beta$ in the TASEP. This involves the sum of squared weights of all configurations with $N$ sites. In the matrix product formalism, the sum takes the form of a tensor product,
\begin{equation}\label{eq:SumSq}
\sum_{C\in\mathcal{L}_N}\mathcal{W}({\cal C})^2 = \bra{W}\otimes \bra{W}(D\otimes D +E\otimes E) ^N\ket{V} \otimes \ket{V} \;.
\end{equation}
For small system sizes, this is readily calculated using the usual reduction relations \eref{eq:Reduction}. For $N = 0, 1, 2$ we calculate these explicitly
\begin{eqnarray}
\fl \bra{W}\otimes \bra{W}(D\otimes D +E\otimes E)^0\ket{V}\otimes\ket{V} = 1 \label{eq:seriesN0} \\
\fl \bra{W}\otimes \bra{W}(D\otimes D +E\otimes E)\ket{V}\otimes\ket{V} = \frac{1}{\beta^2} + \frac{1}{\alpha^2} \label{eq:seriesN1} \\
\fl \bra{W}\otimes \bra{W}(D\otimes D +E\otimes E)^2\ket{V} \otimes \ket{V} \nonumber  \\
\fl =  \bra{W}\otimes \bra{W}(DD\otimes DD+ EE\otimes EE + ED\otimes ED +DE\otimes DE )\ket{V} \otimes \ket{V} \nonumber \\
\fl =  \bra{W}\otimes \bra{W}(DD\otimes DD  + EE\otimes EE + ED\otimes ED \nonumber \\
+D\otimes D + E\otimes E + D\otimes E + E\otimes D)\ket{V} \otimes \ket{V} \nonumber \\
\fl = \frac{1}{\beta^4}+\frac{1}{\alpha^4} + \frac{1}{\alpha^2\beta^2} + \frac{1}{\beta^2} + \frac{1}{\alpha^2} + \frac{2}{\alpha\beta} \label{eq:seriesN2}.
\end{eqnarray}

The problem to be studied is to generalise these expressions to arbitrary $N$. However, using the reduction relations these rapidly become intractable. We present a calculation of the \emph{generating function} of the sum of squared weights, as a function of the reservoir parameters $\alpha$, $\beta$, and a counting parameter $z$ that tracks system size $N$
\begin{equation} \label{eq:Qazb0}
\mathcal{Q}(z;\alpha,\beta) = \sum_{N\geq 0} z^N\bra{W}\otimes \bra{W}(D\otimes D +E\otimes E) ^N\ket{V} \otimes \ket{V}.
\end{equation}
With this, we extract the asymptotic scaling of the sum of squared weights (coefficients of the series expansion in $z$). This will tell us how $H_2$, the entropy, and $\e^{H_2}$, the effective number, scale. As we show in Section~\ref{S:asymp}, singularities in the generating function lead to a different scaling in the three different phases.

The generating function $\mathcal{Q}(z;\alpha,\beta)$ itself is found by interpreting the tensor product expressions in \eref{eq:SumSq} and \eref{eq:Qazb0} as random walks on a lattice. We illustrate the central ideas by considering first of all the simpler problem of calculating the normalisation \eref{eq:ZN} which has previously been obtained by a variety of other means \cite{Blythe2007}.

\section{Mapping to a lattice walk}
\label{S:mapsetup}

One of the known explicit representations of the matrices $D$ and $E$ satisfying \eref{eq:Reduction} involves the ladder operators $g$, $g^\dagger$ that raise and lower state kets $\ket{k}$ (where $g\ket{0} = 0$) \cite{Derrida1993a,Blythe2007}. Specifically,
\begin{eqnarray} 
\label{eq:1drecurrence}
D = \textbf{1} + g & \qquad E = \textbf{1} + g^\dagger \nonumber\\
D\ket{k}=\ket{k}+\ket{k-1} & \qquad E\ket{k}=\ket{k}+\ket{k+1} \nonumber\\
\bra{k} D=\bra{k}+\bra{k+1} & \qquad \bra{k} E=\bra{k}+\bra{k-1} \;. 
\end{eqnarray}
In this representation, the parameters $\alpha$, $\beta$ appear only in the vectors $\bra{W}$, $\ket{V}$. Defining
\begin{eqnarray}
a = \frac{1-\alpha}{\alpha}\;, &\qquad b = \frac{1-\beta}{\beta} \;,
\end{eqnarray}
the boundary vector components are $\braket{W}{m} = (1-ab)a^{m}$, $\braket{m}{V} = b^m$, from which we see that $\braket{W}{V}=1$ as required. One way to write the normalisation $Z_N(\alpha,\beta)$, Eq.~\eref{eq:ZN}, is then
\begin{equation}
\label{eq:normalisation}
Z_N(\alpha,\beta) = (1-ab)\sum_{i\geq 0}\sum_{k \geq 0} a^ib^k\bra{i}\left(g + g^\dagger + 2\cdot\textbf{1}\right)^N\ket{k} \;.
\end{equation}

\begin{figure}
\centering
\label{F:1Dpathwalk}
\includegraphics[scale=.9]{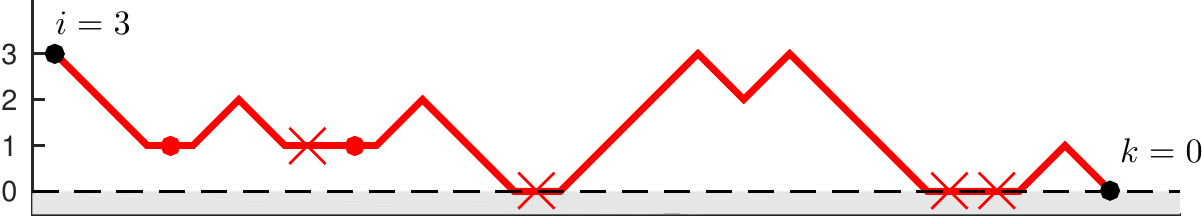}
\caption{Example of the one-dimensional walk, with the two non-movement steps. This walk begins at $i = 3$ and terminates at $k = 0$. The walk can not move below the boundary at $0$.}
\end{figure}

The connection to a lattice walk is to interpret \eref{eq:1drecurrence} as possible \emph{coordinate} changes in a one-dimensional path. Acting on a bra $\bra{i}$, $D$ creates either a step $i\to i+1$ up or a non-movement $i \to i$, denoted $(\uparrow)$, $(\cdot)$ respectively,  and an $E$ imposes a down step $i \to i-1$, or a non-movement $i \to i$, denoted $(\uparrow)$, $(\times)$. (Acting on a ket $\ket{k}$, the directions of the steps are reversed). In this interpretation, the element $\bra{i}\left(g + g^\dagger + 2\cdot\textbf{1}\right)^N\ket{k}$ that appears in \eref{eq:normalisation} counts the number of unique walks of length $N$ comprising the steps $\{\uparrow, \downarrow, \cdot, \times\}$ (where $\cdot$ and $\times$ indicate the two non-movement steps) that start at $i$ and end at $k$, remaining in the upper-half plane (the coordinate $-1$ is an absorbing boundary). See Figure~\ref{F:1Dpathwalk} for an example of such a walk. The normalisation \eref{eq:normalisation} is then a generating function over such paths with all possible combinations of start and end points.

\subsection{Generating function for the sum of weights}
\label{SS:normGF}

One way to obtain the normalisation \eref{eq:normalisation} is to define its generating function
\begin{eqnarray}
\mathcal{Z}(z;a,b) &\equiv \sum_{N \geq 0} z^N\bra{W}(D+E)^N\ket{V} \nonumber \\
\label{eq:normgenfun2}
&= \sum_{N\geq 0}z^N (1-ab)\sum_{i\geq 0}\sum_{k \geq 0} a^ib^k\bra{i}\left(g + g^\dagger + 2\cdot\textbf{1}\right)^N\ket{k} \;,
\end{eqnarray}
which in turn can be calculated using the \emph{kernel method} as we now describe (see \cite{Prodinger2004} for details and further examples).

We consider the $k^{\rm th}$ component of $\mathcal{Z}$
\begin{equation}
\label{eq:muk}
\mu_k(z;a) \equiv \sum_{N\geq 0}z^N \sum_{i\geq 0} a^i\bra{i}\left(g + g^\dagger + 2\cdot\textbf{1}\right)^N\ket{k}
\end{equation}
and obtain a recursion by applying the operator $g + g^\dagger + 2\cdot\textbf{1}$ to the bra $\bra{i}$. Following the definitions in \eref{eq:1drecurrence}, we find
\begin{equation}
\mu_k(z;a) = a^k + z(2+a+\bar{a})\mu_k(z;a) - z\bar{a}\mu_k(z;0) 
\end{equation}
where we have introduced the notation $\bar{a} = 1/a$ that will be used throughout this work. Note that the $a^k$ term arises from the $N=0$ term of \eref{eq:muk}. Rearranging,
\begin{equation}
\label{eq:mukern}
\mu_k(z;a) = \frac{z\mu_k(z;0)-a^{k+1}}{z\left(a-A_-(z)\right)\left(a-A_+(z)\right)}
\end{equation}
where we have factorised the denominator in the variable $a$. We refer to the denominator as the \emph{kernel}. The roots $A_{\pm}(z)$ of the kernel are functions of $z$
\begin{equation}
A_{\pm}(z) = \frac{1-2z\pm\sqrt{1-4z}}{2z}
\end{equation}
with $A_-(z)A_+(z) = 1$. Thus \eref{eq:mukern} exhibits a priori two
poles at $a = A_\pm(z)$. However, as we now argue, one of these poles must be cancelled by the numerator which furnishes the condition that fixes the undetermined function $\mu_k(z;0)$ in \eref{eq:mukern}.

From \eref{eq:muk}, we see that $\mu_k(z;a)$ is a series with non-negative powers of $z$ and $a$. Looking at the denominator of \eref{eq:mukern}, we see that since $A_{+}(z) \to 1/z$ as $z\to0$, a Taylor expansion of this factor about $z=0$ and $a=0$ yields non-negative powers. However, $A_{-}(z)\to0$ as $z\to0$, which generates a spurious $1/a$ term. Since $\mu_k(z;0)$ depends on $z$ (and not on $a$), the only way to eliminate this divergence is to cancel the pole $(a-A_-(z))$ when $a \to z$. This condition fixes $\mu_k(z;0) = A_-(z)^{k+1}/z$ and gives our closed form expression for $\mu_k(z;a)$
\begin{equation}
\mu_k(z;a) = \frac{A_-(z)^{k+1}-a^{k+1}}{z\left(a-A_-(z)\right)\left(a-A_+(z)\right)}. \end{equation}
The full generating function \eref{eq:normgenfun2} is calculable as a  geometric series
\begin{eqnarray}
\mathcal{Z}(z;a,b) &= (1-ab)\sum_{k \geq 0}\left(\frac{A_-(z)^{k+1}-a^{k+1}}{z\left(a-A_-(z)\right)\left(a-A_+(z)\right)}\right)b^k \nonumber \\
&= \frac{1}{z (a-A_+(z)) (bA_-(z)-1)} \;. \label{eq:normGFresultb}
\end{eqnarray}
By introducing $\eta(z) = \frac{1}{2}(1-\sqrt{1-4z}) = z + z^2 + 2z^3 + 5z^4 + \mathcal{O}(z^5)$, which  (up to a factor of $z$) is the generating function for the Catalan numbers $\{C_n\} = \{1,1,2,5,14,42\dots\}$, we have
\begin{equation}
z= \eta(1-\eta),\qquad A_+ = \frac{1-\eta}{\eta},\qquad
A_- = \frac{\eta}{1-\eta} \;,
\end{equation}
and \eref{eq:normGFresultb} can be expressed in a form manifestly symmetric in $(a,b)$
\begin{eqnarray}
\mathcal{Z}(z;a,b) = \frac{1}{\left[1-(1+a)\eta(z)\right]\left[1-(1+b)\eta(z)\right]}\;. \label{eq:normGFresult2}
\end{eqnarray}
Expanding expression \eref{eq:normGFresult2} as a series in $z$ then yields the exact expression \eref{ZNexact}. Alternatively one can sum \eref{ZNexact} and show that \eref{eq:normGFresult2} is obtained (see e.g.~\cite{Blythe2004}).

It is worth noting that the geometric series in \eref{eq:normGFresultb} will have a finite radius of convergence. We perform the calculation assuming we are within this radius of convergence, and extend the domain of the resulting generating function to the full phase diagram (all values of $a>-1$, $b > -1$ of the TASEP) by analytic continuation.

For future reference, it is worth recalling how to extract the leading order behaviour of $Z_N(\alpha,\beta)$  from the nonanalyticity of the generating function that is closest to the origin in the complex-$z$ plane \cite{Wilf2006}. In the simple case of \eref{eq:normGFresult2} we can read these off (see Section \ref{S:asymp}):
\begin{itemize}
\item $\alpha < \frac{1}{2}$, $\alpha < \beta$. A simple pole at $\alpha = \eta(z) \Rightarrow z = \alpha(1-\alpha)$ is dominant. Thus $Z_N(\alpha,\beta) \propto \left(\alpha(1-\alpha)\right)^{-N}$. 
\item $\beta < \frac{1}{2}$, $\alpha > \beta$. A simple pole at $\beta = \eta(z) \Rightarrow z = \beta(1-\beta)$ is dominant. Thus $Z_N(\alpha,\beta) \propto \left(\beta(1-\beta)\right)^{-N}$. 
\item $\alpha > \frac{1}{2}$, $\beta > \frac{1}{2}$. A branch point  at $z = 1/4$ coming from a square root is dominant. Thus $Z_N(\alpha,\beta) \propto 4^N/N^{3/2}$.
\end{itemize}
The dominant singularity of \eref{eq:normGFresult2} changes on lines that  coincide with lines on the phase diagram of Figure~\ref{F:PhaseDia}. We identify phase transitions in the TASEP by the switching of the dominant singularity in the generating function of the normalisation.

\subsection{Generating function for the sum of squared weights}

\begin{figure}
\centering
\includegraphics[scale=.9]{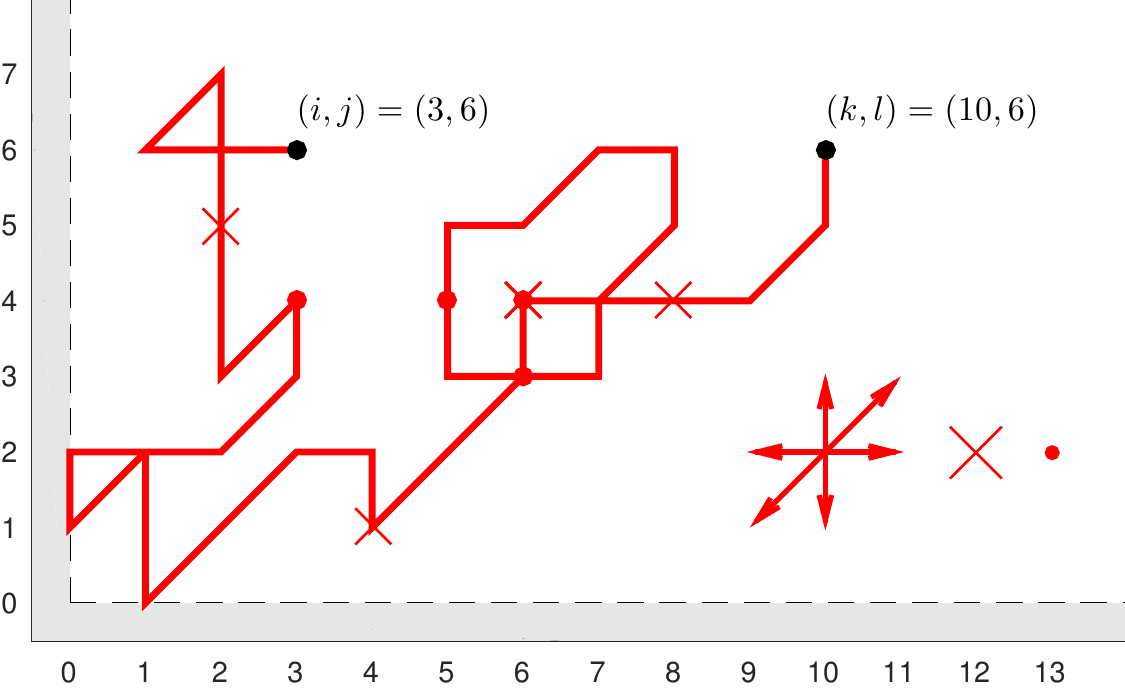}
\caption{Example of the six step walk, with the two non-movement steps, and generalised start and end coordinates. The walk is not self-avoiding and can retrace its own steps. The walk may move along the edges of the boundary, but not beyond.}
\label{fig:sixpathwalk}
\end{figure}

We now turn to the tensor expression in \eref{eq:Qazb0} for the sum of squared weights. Here,
\begin{eqnarray} 
\fl\bra{i}\otimes\bra{j}\left(D\otimes D\right) &=\bra{i}\otimes\bra{j} + \bra{i}\otimes\bra{j+1} + \bra{i+1}\otimes\bra{j} + \bra{i+1}\otimes\bra{j+1} \\
\fl\bra{i}\otimes\bra{j}\left(E\otimes E\right)&=\bra{i}\otimes\bra{j} + \bra{i}\otimes\bra{j-1}+\bra{i-1}\otimes\bra{j} + \bra{i-1}\otimes\bra{j-1} \;.
\end{eqnarray}
These correspond to possible steps of a walk on a two-dimensional lattice spanned by the coordinates  $i$ and $j$. In this interpretation, the summation in \eref{eq:SumSq} is equivalent to the number of unique paths comprising $N$ steps each chosen from the 8 possibilities $\{\nearrow, \rightarrow, \uparrow, \swarrow, \leftarrow, \downarrow, \cdot, \times\}$, where the last two cause no change of position, and where the walk remains in the upper quarter plane. See Figure~\ref{fig:sixpathwalk}. We find
\begin{equation}
\fl \sum_{C\in\mathcal{L}_N}\mathcal{W}({\cal C})^2 = (1-ab)^2\sum_{i \geq 0}\sum_{j\geq 0}\sum_{k\geq 0}\sum_{\ell\geq 0} a^ia^jb^kb^{\ell}\bra{i}\otimes\bra{j}(\mathcal{T}+2\cdot\textbf{1})^N\ket{k}\otimes\ket{\ell} 
\end{equation}
where $\mathcal{T}$ denotes the sum over the tensor operators that correspond to the steps $\{\nearrow, \rightarrow, \uparrow, \swarrow, \leftarrow, \downarrow\}$.

It is helpful to make a change of variable that eliminates the two non-movement steps ($\cdot$ and $\times$), leaving only the six coordinate-changing steps in $\mathcal{T}$. To this end, we define a generating function $\mathcal{R}$, that counts the number of walks comprising $N$ steps from $\mathcal{T}$, that begin at $(i,j)$ and end at $(k,\ell)$, remaining in the upper quarter plane
\begin{equation} 
\label{eq:Rxtyvw}
\fl \mathcal{R}(t;x,y,v,w) = \sum_{N \geq 0}\sum_{i\geq 0}\sum_{j\geq 0}\sum_{k\geq 0}\sum_{l\geq 0} t^Nx^iy^jv^kw^l\bra{i}\otimes\bra{j}\mathcal{T}^N\ket{k}\otimes\ket{\ell} \;.
\end{equation}
The  five variables $(t,x,y,v,w)$ of $\mathcal{R}$  correspond to fugacities for the path length $N$, its start coordinates $(i,j)$ and its end coordinates $(k,\ell)$, respectively.  To relate $\mathcal{Q}$ given by \eref{eq:Qazb0} to $\mathcal{R}$, we use the identity
\begin{equation}
\sum_{N\ge0} z^N (X+Y)^N = \sum_{N\ge0} \sum_{P\ge0} {N \choose P} (zX)^P (zY)^{N-P} = \sum_{P\ge0} \frac{(zX)^P}{(1-zY)^{P+1}} 
\end{equation}
for commuting objects $X$ and $Y$. Then,
\begin{eqnarray} 
\fl \mathcal{Q}(z;\alpha,\beta) &=\sum_{N \geq 0}\sum_{i\geq 0}\sum_{j\geq 0}\sum_{k\geq 0}\sum_{\ell\geq 0} z^N(1-ab)^2a^ia^jb^kb^{\ell}\bra{i}\otimes\bra{j}(\mathcal{T}+2\cdot\textbf{1})^N\ket{k}\otimes\ket{\ell} \nonumber  \\
\fl &= \frac{(1-ab)^2}{1-2z}\sum_{i\geq 0}\sum_{j\geq 0}\sum_{k\geq 0}\sum_{\ell\geq 0} a^{i+j}b^{k+\ell}\sum_{P\geq 0} \left(\frac{z}{1-2z}\right)^P\bra{i}\otimes\bra{j}\mathcal{T}^P\ket{k}\otimes\ket{\ell}  \nonumber  \\
\fl &= \frac{(1-ab)^2}{1-2z}\mathcal{R}\left(\frac{z}{1-2z};\frac{1-\alpha}{\alpha},\frac{1-\alpha}{\alpha},\frac{1-\beta}{\beta},\frac{1-\beta}{\beta}\right) \label{eq:abtrans}
. 
\end{eqnarray}
Thus $\mathcal{R}$ and $\mathcal{Q}$ are related by the transformation $t = z/(1-2z)$, $z \in \left[0,\frac{1}{2}\right)$. We focus on finding an expression for $\mathcal{R}$, generalising the result of \cite{Bousquet2010} where the end point of this six-step walk was fixed at the origin, $k=\ell=0$.

\subsection{Recurrence relation}
We now state the recurrence relation that $\mathcal{R}(t;x,y,v,w)$ obeys. If we apply the leftmost $\mathcal{T}$ operator in \eref{eq:Rxtyvw} to the bra $\bra{i}\otimes\bra{j}$,
\begin{eqnarray} 
\label{eq:recurrence}
\fl \mathcal{R}(t;x,y,v,w) = & \sum_{i\geq 0}\sum_{j\geq 0}\sum_{k\geq 0}\sum_{l\geq 0} x^iy^jv^kw^l\delta_{ik}\delta_{jl} \\
&+ \sum_{i\geq 0}\sum_{j\geq 0}\sum_{k\geq 0}\sum_{l\geq 0}\sum_{N\geq 1} t^Nx^iy^jv^kw^l\bigg[\bra{i+1}\bra{j} + \bra{i-1}\bra{j} + \bra{i}\bra{j+1}  \nonumber \\ 
& + \bra{i}\bra{j-1} + \bra{i+1}\bra{j+1} + \bra{i-1}\bra{j-1}\bigg]\mathcal{T}^{(N-1)}\ket{k}\ket{\ell}  \nonumber
\end{eqnarray}
where we suppress the outer product $\otimes$ symbol to lighten the notation. The first term is the $N=0$ contribution, and the second $N\geq 1$ term makes explicit the six possible steps the walk can make. By the definition of $\mathcal{R}$, \eref{eq:recurrence} is rewritten
\begin{eqnarray}
\label{eq:Pxy0a}
\fl \mathcal{R}(t;x,y,v,w) = & \frac{1}{(1-xv)(1-yw)}  + t\bar{x}\bar{y}\mathcal{R}(t;0,0,v,w)  \\
& - t(\bar{x}\bar{y} + \bar{x})\mathcal{R}(t;0,y,v,w) - t(\bar{x}\bar{y} + \bar{y})\mathcal{R}(t;x,0,v,w)  \nonumber\\
& +t(x+\bar{x}+y+\bar{y}+xy+\bar{x}\bar{y})\mathcal{R}(t;x,y,v,w) \;,  \nonumber
\end{eqnarray}
recalling that $\bar{x}=1/x$ and $\bar{y}=1/y$. At this point we define the \emph{kernel} $K(x,y,t)$ and introduce a shorthand $K_{xy}$
\begin{eqnarray} \label{eq:kernel}
K(x,y,t) \equiv K_{xy} = 1 - t(x+\bar{x}+y+\bar{y}+xy+\bar{x}\bar{y})
\end{eqnarray}
and \eref{eq:Pxy0a} becomes
\begin{eqnarray}
\label{eq:mainrecurrence}
\fl K_{xy}\mathcal{R}(t;x,y,v,w)  =& \frac{1}{(1-xv)(1-yw)}  + t\bar{x}\bar{y}\mathcal{R}(t;0,0,v,w) \\
& - t(\bar{x}\bar{y} + \bar{x})\mathcal{R}(t;0,y,v,w) - t(\bar{x}\bar{y} + \bar{y})\mathcal{R}(t;x,0,v,w) \nonumber.
\end{eqnarray}
By making the substitution $(x,y,v,w) \rightarrow (a,a,b,b)$, we find a simplification using the symmetry of the walk
\begin{eqnarray}
\label{eq:simplerecurrence}
\fl a^2K_{aa}\mathcal{R}(t;a,a,b,b) = \frac{a^2}{(1-ab)^2} + t\mathcal{R}(t;0,0,b,b) - 2t(1+a)\mathcal{R}(t;a,0,b,b).
\end{eqnarray}
We see from \eref{eq:simplerecurrence} that an expression for $\mathcal{R}(t;a,0,b,b)$ is sufficient to find $\mathcal{R}(t;a,a,b,b)$, the generating function for the random walk, and in turn $\mathcal{Q}(z;\alpha,\beta)$, the generating function for the TASEP weights. However, therein lies the difficulty in this problem. With the recursion relation in \eref{eq:mainrecurrence} we encounter an \emph{obstinate kernel}. With the kernel as a function of two variables $x$, $y$, there is insufficient information in the single recurrence relation to fix the right hand side terms of \eref{eq:mainrecurrence} by a pole-cancelling method, as used in the case of the normalisation in Section~\ref{SS:normGF}. Instead, we must turn to the more sophisticated method of \cite{Bousquet2005,Bousquet2010} which exploits a symmetry property of the kernel to solve for $\mathcal{R}$.

\subsection{Factorisation of the kernel}

The kernel $K_{xy}$ \eref{eq:kernel} is central to this calculation. We define its two roots in the variable $y$ as $Y_-(x,t)$, $Y_+(x,t)$ and write it in factorised form
\begin{eqnarray}
K_{xy} &= -\frac{t(1+x)}{y}\left(y-Y_-(x,t)\right)\left(y-Y_+(x,t)\right) \label{eq:Kxyfact} \\
Y_{\pm}(x,t) &= \frac{1-t(\bar{x}+x)\pm\sqrt{\Delta(x,t)}}{2t(1+x)} \label{eq:Y0Y1}
 \\
\Delta(x,t) &= 1-6t^2 + t^2(x^2+\bar{x}^2) -2t(1+2t)(x+\bar{x})  
\end{eqnarray}
where $\Delta(x,t)$ is termed the \emph{discriminant}. One can show that $Y_-(x,t)Y_+(x,t) = \bar{x}$. Knowing these roots, we rewrite the kernel in a form we later use
\begin{equation}
\label{eq:factorisedkernel}
\frac{1}{K(x,y,t)} = \frac{1}{\sqrt{\Delta(x,t)}}\left[\frac{1}{1-\bar{y}Y_-(x,t)}+\frac{1}{1-y\bar{Y}_+(x,t)}-1\right].
\end{equation}
The discriminant can also be factorised, and we write it as a product of its roots
\begin{eqnarray}
\label{eq:discriminant}
\Delta(x,t) &= \Delta_0\Delta_+(x,t)\Delta_+(\bar{x},t) \\
\Delta_0(t) &= \frac{t^2}{X_-(t)X_+(t)} \\
\Delta_+(x,t) &= (1 - X_-(t) x)(1 - X_+(t) x).
\end{eqnarray}
The roots $X_-(t)$, $X_+(t)$ are functions of $t$ alone
\begin{eqnarray}
\label{eq:XpXm}
X_{\pm}(t)
= \frac{2 t+1}{2 t} \pm \sqrt{\frac{3t+1}{t}}-\frac{\sqrt{2 t+1}}{2 t}\sqrt{1+6t\pm4\sqrt{t(3t+1)}}.
\end{eqnarray}
We refer to these expressions throughout the calculation. We also define the following transformations here, that we use on applying \eref{eq:abtrans} 
\begin{eqnarray}\label{eq:zidentities}
\fl \Lambda_{\pm}(z) & = 2zX_{\pm}\left(\frac{z}{1-2z}\right) = 1\pm 2\sqrt{z(1+z)}-2\sqrt{z+\frac{1}{4}\pm\sqrt{z(1+z)}} \\
\fl \Gamma(\alpha,z) & \equiv \Gamma(\alpha) = \Delta_+\left(\frac{1-\alpha}{\alpha},\frac{z}{1-2z}\right) = \left[1-\frac{1-\alpha}{2z\alpha}\Lambda_-(z)\right]\left[1-\frac{1-\alpha}{2z\alpha}\Lambda_+(z)\right].
\end{eqnarray}

\section{Generating function for the $\alpha = \beta = 1$ weights, $\mathcal{Q}(z;1,1)$}
\label{S:Qz11}

In this section we present a full calculation of  $\mathcal{Q}(z;1,1)$ for the case $\alpha = \beta =1$ where expressions simplify considerably from the case of general $\alpha$ and $\beta$. We find that with increasing generality of the generating function, the algebra becomes more elaborate, but the principles remain the same. By working through this restricted case in detail we aim to clearly outline this method, while the algebra is  simple compared to the general $\alpha,\beta$ case which is deferred to an appendix. In this simpler case, our solution follows closely the method of \cite{Bousquet2010}.

For brevity, define
\begin{equation}
\mathcal{P}(x,y) \equiv \mathcal{R}(t;x,y,0,0)
\end{equation}
whereby we now have any functional $t$ dependence as implicit. Because we have set $v = w = 0$ (the equivalent of fixing $\beta = 1$ in the TASEP) the recurrence relation \eref{eq:mainrecurrence} reduces to  
\begin{equation}
\label{eq:Pxy}
\fl xy\mathcal{P}(x,y) = \frac{1}{K_{xy}}\left[xy - t(1+x)\mathcal{P}(x,0) - t(1+y)\mathcal{P}(0,y) + t\mathcal{P}(0,0)\right].
\end{equation}
It is at this point we use an important property of the kernel
\begin{equation} \label{eq:kersym}
K(x,y,t) = K(x,\bar{x}\bar{y},t) = K(\bar{x}\bar{y},y,t)
\end{equation}
to acquire two additional expressions from \eref{eq:Pxy}
\numparts
\begin{eqnarray}
\fl \bar{y}\mathcal{P}(x,\bar{x}\bar{y}) &= \frac{1}{K_{xy}}\left[\bar{y} - t(1+x)\mathcal{P}(x,0) - t(1+\bar{x}\bar{y})\mathcal{P}(0,\bar{x}\bar{y}) + t\mathcal{P}(0,0)\right]\label{eq:a2} \\ 
\fl \bar{x}\mathcal{P}(\bar{x}\bar{y},y) &= \frac{1}{K_{xy}}\left[\bar{x} - t(1+\bar{x}\bar{y})\mathcal{P}(\bar{x}\bar{y},0) - t(1+y)\mathcal{P}(0,y) + t\mathcal{P}(0,0)\right] \label{eq:a3}.
\end{eqnarray}
\endnumparts 
These can be combined as $\left(\eref{eq:Pxy} + \eref{eq:a2} - \eref{eq:a3}\right)$ to give
\begin{eqnarray} 
\label{eq:Pxy2}
\fl xy\mathcal{P}(x,y) + \bar{y}\mathcal{P}(x,\bar{x}\bar{y}) - \bar{x}\mathcal{P}(\bar{x}\bar{y},y) \\ 
= \frac{1}{K_{xy}}\left[t\mathcal{P}(0,0) - 2t(1+x)\mathcal{P}(x,0) + xy + \bar{y} - \bar{x} \right]. \nonumber
\end{eqnarray}
The exploitation of this kernel symmetry is the key step in solving an otherwise insufficient recurrence relation \eref{eq:Pxy}. Crucially, we are now able to find closed form expressions for the generating functions $\mathcal{P}(0,0)$, $\mathcal{P}(x,0)$ by extracting coefficients of certain powers of $x$ and $y$. This is because we have used the kernel symmetry to make nearly all $y$-dependence in expression \eref{eq:Pxy2} explicit. 

With this in mind, we rewrite \eref{eq:Pxy2} with the factorised kernel  \eref{eq:factorisedkernel}
\begin{eqnarray} 
\label{eq:Pxy3}
\fl xy\mathcal{P}(x,y) + \bar{y}\mathcal{P}(x,\bar{x}\bar{y}) - \bar{x}\mathcal{P}(\bar{x}\bar{y},y) \\
\fl=  \frac{1}{\sqrt{\Delta(x)}}\left[\frac{1}{1-\bar{y}Y_-(x)}+\frac{1}{1-y\bar{Y}_+(x)}-1\right]\left[t\mathcal{P}(0,0) - 2t(1+x)\mathcal{P}(x,0) + xy + \bar{y} - \bar{x} \right] \nonumber.
\end{eqnarray}

We first want $\mathcal{P}(0,0)$. Knowing that this is a function of $t$ alone, we need the $x^0y^0$ coefficient of \eref{eq:Pxy3}. Having made most of the $y$-dependence explicit, we begin by extracting the $y^0$ component.

\subsection{$y^0$ coefficient extraction}

By making explicit the power series on the left hand side (LHS) of \eref{eq:Pxy3}
\begin{eqnarray}
\fl \mathrm{LHS}\ \eref{eq:Pxy3} = \sum_{i\geq 0}\sum_{j\geq 0}\sum_{N\geq 0} \left[x^{i+1}y^{j+1} + x^{i-j}y^{-j-1} - x^{-i-1}y^{j-i}\right]t^N\bra{i}\bra{j}\mathcal{T}^N\ket{0}\ket{0}
\end{eqnarray}
we read off the $y^0$ component, $\{y^0\}$, as 
\begin{equation}
\left\{y^0\right\} \mathrm{LHS} \ \eref{eq:Pxy3} = - \bar{x}\sum_{i\geq 0}\sum_{N \geq 0} \bar{x}^{i}t^N\bra{i}\bra{i}\mathcal{T}^N\ket{0}\ket{0} \equiv -\bar{x}P_D(\bar{x})
\end{equation}
where $\mathcal{P}(x,x) \equiv \mathcal{P}_D(x)$ is the generating function for walks comprising steps from $\mathcal{T} = \left\{\nearrow, \rightarrow, \uparrow, \swarrow, \leftarrow, \downarrow\right\}$, from the origin, remaining in the upper quarter plane and  terminating on the diagonal. We now express the RHS of \eref{eq:Pxy3} as a formal power series
\begin{eqnarray}
\fl \mathrm{RHS} \ \eref{eq:Pxy3} \\
\fl = \frac{1}{\sqrt{\Delta(x)}}\left[\sum_{i\geq 0}y^{-i}Y_-(x)^i + \sum_{j\geq 0}y^jY_+(x)^{-j} - 1 \right]\left[t\mathcal{P}(0,0) - 2t(1+x)\mathcal{P}(x,0) + xy + \bar{y} - \bar{x}\right] \nonumber
\end{eqnarray}
and read off the $y^0$ component to leave
\begin{equation} 
\label{eq:Pxy4}
-\bar{x}\mathcal{P}_D(\bar{x}) = \frac{1}{\sqrt{\Delta(x)}}\left[t\mathcal{P}(0,0)-2t(1+x)\mathcal{P}(x,0)-\bar{x}+2xY_-(x)\right]
\end{equation}
having used $Y_+(x)Y_-(x) = \bar{x}$. From \eref{eq:Pxy4}, and the factorisation \eref{eq:discriminant} it is a simple matter to  determine $\mathcal{P}(0,0)$.

\subsection{$x^0$ coefficient extraction, $\mathcal{Q}(z;1,1)$ result}

Inserting the explicit form \eref{eq:Y0Y1} for $Y_-$ and factorising the discriminant \eref{eq:discriminant}, we can rearrange the terms in expression \eref{eq:Pxy4} into the form
\begin{eqnarray}
\label{eq:Pxy5}
\fl \sqrt{\Delta_+(\bar{x})}\left[\frac{x}{t}-(1+\bar{x})\mathcal{P}_D(\bar{x})\right] \\ 
\fl = \frac{1}{\sqrt{\Delta_0\Delta_+(x)}}\left[(1+x)t\mathcal{P}(0,0)-2t(1+x)^2\mathcal{P}(x,0)+\frac{x}{t}-2-\bar{x}-x^2\right]\nonumber.
\end{eqnarray}
We now perform two expansions with the factorised form of the discriminant \eref{eq:discriminant}
\begin{eqnarray}
\sqrt{\Delta_+(\bar{x})} &= 1 - \frac{1}{2}(X_-+X_+)\bar{x} + \mathcal{O}(\bar{x}^2) \\
\frac{1}{\sqrt{\Delta_+(x)}} &= 1 + \frac{1}{2}(X_-+X_+)x + \mathcal{O}(x^2)
\end{eqnarray}
and extract the $x^0$ component of \eref{eq:Pxy5} to find
\begin{eqnarray} 
\mathcal{R}(t;0,0,0,0) &= \mathcal{P}(0,0) \nonumber \\ 
&= \frac{1}{2t}\frac{(X_-+X_+)-\sqrt{X_-X_+}\left(4+X_-+X_+\right)}{\sqrt{X_-X_+}-1} \nonumber \\
& = 1 + 3t^2 + 4t^3 + 26t^4 + \mathcal{O}(t^5) \label{eq:P00t}
\end{eqnarray}
in terms of the roots of the discriminant $X_-(t)$, $X_+(t)$ \eref{eq:XpXm}. This is a verification of \cite{Bousquet2010}. To recall, $\mathcal{R}(t;0,0,0,0)$ generates the numbers of walks of $N$ steps from $\{\nearrow, \rightarrow, \uparrow, \swarrow, \leftarrow, \downarrow\}$ in the upper quarter plane that  start and finish at the origin. To find the generating function for the sum of squared weights in a TASEP of length $N$, we apply \eref{eq:abtrans}
\begin{eqnarray}
\mathcal{Q}(z;1,1) = \frac{1}{1-2z}\mathcal{R}\left(\frac{z}{1-2z};0,0,0,0\right)
\label{eq:Qz11trans}
\end{eqnarray}
to acquire a preliminary expression, which we simplify by denesting the square roots. We make extensive use of the identity 
\begin{equation}
\label{eq:denesting}
\sqrt{2}\sqrt{A+B\sqrt{C}} = \sqrt{A+\sqrt{A^2-B^2C}} + \sqrt{A-\sqrt{A^2-B^2C}}
\end{equation}
which reduces a nested square root into a sum of two square roots, if $A^2-B^2C$ is a perfect square. Using this, we eventually find \eref{eq:Qz11trans} in the simplest form to be
\begin{eqnarray}
\fl \mathcal{Q}(z;1,1) \nonumber \\
\fl\quad = \frac{1}{4z^2}\left[3\sqrt{2z}\sqrt{1-2z-\sqrt{1-8z}} +\sqrt{2(1+z)}\sqrt{1-2z+\sqrt{1-8z}}-4z-2\right] \;.
\label{eq:Qz11result}
\end{eqnarray}
Note that the first double square-root expression vanishes as $z\to0$, indicating that this generating function is a series in positive powers of $z$, as required. Expanding about the origin, we find
\begin{eqnarray}
\mathcal{Q}(z;1,1) = 1 + 2z + 7z^2 + 30z^3 + 146z^4 + 772z^5 + 4331z^6 + \mathcal{O}(z^7) \;.
\label{eq:Qz11series}
\end{eqnarray}

The expression \eref{eq:Qz11result} is the first key result of this paper. The coefficients in the power series expansion \eref{eq:Qz11series} match with the enumerated sums of squared TASEP weights, in the case $\alpha = \beta = 1$, for systems of size $N = (0,1,2,3,\dots)$. This set of coefficients $\{q_N\} = \{1,2,7,30,146,772,4331,\dots\}$ matches with those of sequence A196148 in the OEIS \cite{oeis}, which take the form 
\begin{eqnarray}
q_N &= \sum_{P=0}^N\frac{(2N+1)!(N+1)!}{(2P+1)!(2N-2P+1)!(P+1)!(N-P+1)!}.
\label{eq:a1b1}
\end{eqnarray}
In fact, expression \eref{eq:a1b1} arises in the literature as the solution to a problem involving the summation of the squared weights, in a combinatorial problem involving \emph{path dominance} \cite{Kreweras1981}. Outside of this work, we have proven a mapping between weights in this problem and weights of the TASEP, confirming the result \eref{eq:a1b1} \cite{ourselves}.

To summarise, we now have the generating function \eref{eq:Qz11result} for the sum of squared weights at the point $\alpha = \beta = 1$ on the phase diagram. With the same method of applying the kernel symmetry in \eref{eq:Pxy2} and extracting coefficients, we now extend this approach to find generating functions first for arbitrary $\alpha$ but $\beta=1$, and subsequently for arbitrary $\alpha$, $\beta$.

\subsection{$x^+$ coefficient extraction, obtaining $\mathcal{Q}(z;\alpha,1)$}
\label{S:Qza1}

Having found $\mathcal{R}(t;0,0,0,0)$, we now generalise to 
$\mathcal{R}(t;a,a,0,0)$. This corresponds to the line $\beta =1$ which traverses the low density and maximal current phases as $\alpha$ is varied. Noticing the recursion relation in \eref{eq:Pxy}, this requires $\mathcal{R}(t;x,0,0,0)$ (by symmetry, $\mathcal{R}(t;0,y,0,0)$ follows). We return to \eref{eq:Pxy3}, and obtain an expression for $\mathcal{P}(x,0)$ by considering the positive powers of $x$. The LHS is elementary
\begin{eqnarray}
\left\{x^+\right\} \mathrm{LHS} \ \eref{eq:Pxy5} & = \frac{x}{t}
\end{eqnarray}
where $\left\{x^+\right\}$ denotes `the positive powers in $x$ within'. The RHS is more involved, and we explicitly subtract any $\mathcal{O}(\bar{x})$
\begin{eqnarray}
\fl \left\{x^+\right\} \mathrm{RHS} \ \eref{eq:Pxy5} =& \frac{1}{\sqrt{\Delta_0\Delta_+(x)}}\Bigg[t(1+x)\mathcal{P}(0,0)-2t(1+x)^2\mathcal{P}(x,0) -2 \\ 
&-x^2-\bar{x}+\frac{x}{t}\Bigg] + \frac{1}{\sqrt{\Delta_0}}\left[t\mathcal{P}(0,0)+2+\bar{x}+\frac{1}{2}\left(X_-+X_+\right)\right] \nonumber \;.
\end{eqnarray}
This gives for $\mathcal{P}(x,0)$
\begin{eqnarray}\label{eq:Px0}
\fl \mathcal{P}(x,0) = -\frac{1}{2t(1+x)^2}\Bigg[\left(\sqrt{\Delta_0\Delta_+(x)}-1\right)\frac{x}{t}+2-t(1+x)\mathcal{P}(0,0) \\
+x^2+\bar{x}-\sqrt{\Delta_+(x)}\left(t\mathcal{P}(0,0)+2+\bar{x}+\frac{1}{2}\left(X_-+X_+\right)\right)\Bigg] \nonumber \;.
\end{eqnarray}

From this, the steps to finding $\mathcal{Q}(z;\alpha,1)$ are straightforward. Using \eref{eq:Pxy}, we acquire $\mathcal{P}(x,y) = \mathcal{R}(t;x,y,0,0)$ from this new result, whereby we find the generating function for squared weights for the general $\alpha$, $\beta = 1$ case after applying  the transformation \eref{eq:abtrans}. With further algebraic manipulation, we eventually find
\begin{eqnarray} \label{eq:alphagenfun}
\fl \mathcal{Q}(z;\alpha,1)  &= \frac{1}{1-2z}\mathcal{P}\left(\frac{z}{1-2z};\frac{1-\alpha}{\alpha},\frac{1-\alpha}{\alpha}\right) \nonumber \\
\fl &= -\frac{\alpha ^2}{z (1-\alpha )}-\frac{\sqrt{\Gamma(\alpha)} \alpha ^5
   \left(1+\frac{1}{1-\alpha }+\frac{2 \sqrt{\Lambda_-\Lambda_+}}{2 z-\sqrt{\Lambda_-
   \Lambda_+}}-\frac{2 z (1-\alpha )}{\sqrt{\Lambda_-\Lambda_+} \alpha }\right)}{\alpha^2(1-\alpha)^2-z\left(\alpha^2+(1-\alpha)^2\right)}  \nonumber \\
\fl &= 1 + \left(1+\frac{1}{\alpha^2}\right)z+\left(2+\frac{2}{\alpha}+\frac{2}{\alpha^2} + \frac{1}{\alpha^4}\right)z^2 + \cdots \;.
\end{eqnarray}
The coefficients of this power series in $z$ are the sums of squared weights of the TASEP for increasing system size with $\beta = 1$, and match with those calculated in  (\ref{eq:seriesN0}-\ref{eq:seriesN2}) using the matrix reduction relations.

By the symmetry between $\alpha$ and $\beta$, we also have from \eref{eq:alphagenfun} $\mathcal{Q}(z;1,\beta)$, whereby $\alpha$ is fixed and $\beta$ is variable. This gives us information along two lines in the phase diagram, crossing at $\alpha = \beta = 1$.

\section{Generating function for the general $\alpha, \beta$ weights, $\mathcal{Q}(z;\alpha,\beta)$}
\label{S:Qzab}

We come at last to the generating function across the whole phase diagram, $\mathcal{Q}(z;\alpha,\beta)$, for which we require an expression for $\mathcal{R}(t;a,a,b,b)$. With a view to brevity, define
\begin{equation}
\mathcal{R}(x,y) \equiv \mathcal{R}(t;x,y,b,b).
\end{equation}
We recall the recurrence relation \eref{eq:mainrecurrence} for this function
\begin{equation}
\label{eq:Rxy0a}
\fl K_{xy}\mathcal{R}(x,y) \\ 
= \frac{1}{(1-bx)(1-by)} + t\bar{x}\bar{y}\mathcal{R}(0,0) - t(\bar{x}\bar{y} + \bar{x})\mathcal{R}(0,y) - t(\bar{x}\bar{y} + \bar{y})\mathcal{R}(x,0) \nonumber.
\end{equation}
This differs in structure from \eref{eq:Pxy} only in the change of the first term, to embed factors of $b = (1-\beta)/\beta$. While this is a different recurrence relation, we use the same approach here as in Section \ref{S:Qz11}: exploiting the symmetry of the kernel to obtain an expression, which we can extract coefficients from to obtain a closed form for the generating function. However, compared to the $\beta = 1$ case of Section \ref{S:Qz11}, these additional factors of $b$ add a surprising degree of algebraic complication to the calculation.

Nonetheless, we use the symmetry property of the kernel \eref{eq:kersym} to arrive at 
\begin{eqnarray} 
\label{eq:Rxy1}
\fl xy\mathcal{R}(x,y) +\bar{y}\mathcal{R}(x,\bar{x}\bar{y}) - \bar{x}\mathcal{R}(\bar{x}\bar{y},y) = \frac{1}{\sqrt{\Delta(x)}}\Bigg[\frac{1}{1-\bar{y}Y_-(x)}+\frac{1}{1-y\bar{Y}_+(x)} -1\Bigg]\times \\ 
\fl \left[\frac{xy}{(1-bx)(1-by)} +\frac{\bar{y}}{(1-bx)(1-b\bar{x}{\bar{y}})} - \frac{\bar{x}}{(1-b\bar{x}\bar{y})(1-by)}+t\mathcal{R}(0,0)-2t(1+x)\mathcal{R}(x,0)\right] \nonumber
\end{eqnarray}
which recovers \eref{eq:Pxy3} in the case $b = 0$. We then extract the $\{y^0\}$, $\{x^+\}$ components of \eref{eq:Rxy1}, to obtain a closed form expression for $\mathcal{R}(t;a,a,b,b)$. This full coefficient extraction is outlined in \ref{S:xextraction}, with the result for $\mathcal{R}(t;a,a,b,b)$ in equation \eref{eq:Raatbbfinal}.

\subsection{$\mathcal{Q}(z;\alpha,\beta)$ result}
\label{S:Qazb}

Having performed this coefficient extraction, we find an expression for $\mathcal{R}(t;a,a,b,b)$, which we quote in the appendix, equation \ref{eq:Raatbbfinal}. With further algebraic manipulation, we obtain the full generating function $\mathcal{Q}(z;\alpha,\beta)$ with the transformation \eref{eq:abtrans} (recalling the definitions in \eref{eq:zidentities})
\begin{eqnarray}
\fl \mathcal{Q}(z;\alpha,\beta) = \frac{(1-ab)^2}{1-2z}\mathcal{R}\left(\frac{z}{1-2z},\frac{1-\alpha}{\alpha},\frac{1-\alpha}{\alpha},\frac{1-\beta}{\beta},\frac{1-\beta}{\beta}\right) \nonumber \\
\fl = -\frac{\alpha ^2 \beta ^2}{z (\alpha +\beta -2 \alpha  \beta )+(\alpha +\beta -1-\alpha  \beta ) \alpha  \beta } + \sqrt{\frac{\Gamma (\alpha)\Gamma (\beta)}{\Lambda_-\Lambda_+}}\times \nonumber \\ 
\fl \left[\frac{ (1-\alpha-\beta ) \alpha \beta  z^2z_0(\alpha)z_0(\beta)}{4(z_0(\alpha)-z)(z_0(\beta)-z)(1-\alpha )^2 (1-\beta )^2(z [\alpha +\beta -2 \alpha  \beta ]+\alpha  \beta  [\alpha +\beta -1-\alpha  \beta ])}\right]\times \nonumber \\
\fl\Bigg[-2 \sqrt{1-8 z} (1-\alpha -\beta )^2 \nonumber  + 8 z (\alpha +\beta -2 \alpha  \beta )-2 (\alpha +\beta -2 \alpha  \beta -1)^2 \\ 
 -(1-\alpha -\beta ) \left(\sqrt{1-8 z}+(1-2 \alpha ) (1-2 \beta )\right)\sqrt{2+8z+2\sqrt{1-8 z}} \Bigg]  \label{eq:Qzab}
\end{eqnarray}
where
\begin{eqnarray}\label{eq:z0}
z_0(\gamma) = \frac{\gamma^2(1-\gamma)^2}{\gamma^2 + (1-\gamma)^2}.
\end{eqnarray}
This is the most  general result of this paper. It would be of no surprise if further simplifications to this generating function were found. However, we only want the asymptotic scaling of the coefficients of the function's power series. For this purpose, \eref{eq:Qzab} is sufficiently simple.

Before analysing \eref{eq:Qzab} in detail, we notice immediately that upon fixing the function along the factorisation line, $\beta = 1-\alpha$, we recover
\begin{equation}
\mathcal{Q}(z;\alpha,1-\alpha) = \frac{1}{1-z(\frac{1}{\alpha^2}+\frac{1}{(1-\alpha)^2})}
\end{equation}
from the first term in \eref{eq:Qzab}; the second term vanishes. This recovers a generating function for the sum of squared weights, for the case discussed in Section \ref{SS:TASEP} of a \emph{Bernoulli} distribution. This serves as one verification of our method. In addition, one can compute the series expansion of $\mathcal{Q}(z;\alpha,\beta)$ in $z$, to verify that its coefficient series is indeed the sums of squared weights for increasing system size, the first few having been directly evaluated in (\ref{eq:seriesN0}-\ref{eq:seriesN2}).

\section{Asymptotic analysis of the generating function}
\label{S:asymp}
To summarise, we now have with \eref{eq:Qzab} a closed form expression for 
\begin{eqnarray}
\mathcal{Q}(z;\alpha,\beta) &= \sum_{N\geq 0} z^N\bra{W}\otimes \bra{W}(D\otimes D +E\otimes E) ^N\ket{V} \otimes \ket{V} \nonumber \\
& = \sum_{N \geq 0}  z^N \left(\sum_{C\in\mathcal{L}_N} \mathcal{W}({\cal C})^2 \right)
\end{eqnarray}
where $\mathcal{W}(C\in\mathcal{L}_N)$ are the weights of TASEP configurations of length N. We use this to find the scaling of the sum of squared probabilities, with a view to finding an expression for the R\'{e}nyi entropy \eref{eq:H2}. We use standard asymptotic methods, following \cite{Wilf2006}. Based on the form of the generating function $\mathcal{Q}(z;\alpha,\beta)$ \eref{eq:Qzab}, we expect poles at $z = z_0(\alpha)$, $z=z_0(\beta)$, and a branch point at $z = 1/8$. We outline the asymptotic method for these two different cases, in order to establish notation.

\subsection{Asymptotic method}

For a generating function $\mathcal{F}(z) = \sum_{N\geq 0} z^Nc_N$, the leading-order asymptotic scaling of $c_N$  is determined by the value of $z$ closest to the origin, $z^*$, such that $\mathcal{F}(z^*)$ is nonanalytic \cite{Wilf2006}. For the case of a simple pole, we perform a series expansion about the pole to acquire 
\begin{equation} 
\label{eq:poleexpand}
\mathcal{F}(z) = \frac{g_{-1}}{z-z^*} + \sum_{j \geq 0}g_{j}(z-z^*)^j 
\end{equation}
whereby the coefficients $c_N$ have the asymptotic form
\begin{equation}
\label{eq:coeffscale1}
c_N \sim -\frac{g_{-1}}{z^*}\left(z^*\right)^{-N} \;.
\end{equation}
In the case of a branch point being the first singularity, a series expansion about this yields an imaginary contribution:
\begin{equation}
\mathcal{F}(z) = ih_{k}(z-z^*)^{k} + \sum_{j\geq 0}h_j(z-z^*)^j
\end{equation}
where $k$ is non-integer and $h_k$ is real. In this case, the coefficients $c_N$ scale as
\begin{equation}
\label{eq:coeffscale2}
c_N \sim \frac{h_k\left(z^*\right)^k}{\Gamma(-k)}N^{-(k+1)}(z^*)^{-N}
\end{equation}
with $\Gamma$ the usual gamma function.

\subsection{Low density phase $\alpha < \beta$, $\alpha < \frac{1}{2}$}

In the low density phase, the first singularity we identify in $\mathcal{Q}(z;\alpha,\beta)$ is a simple pole, at $z = z_0(\alpha)$ \eref{eq:z0}. We find an elaborate expression for $g_{-1}$ presented in \ref{S:Residues} \eref{eq:g1}. As there is an $\alpha$ dependence in the \emph{location} of the pole $z_0(\alpha)$, the value of $\alpha$ affects the R\'{e}nyi entropy to the leading order. 
\begin{figure}
\label{fig:poleresidue}
\includegraphics[trim=0mm 5mm 0mm 0mm]{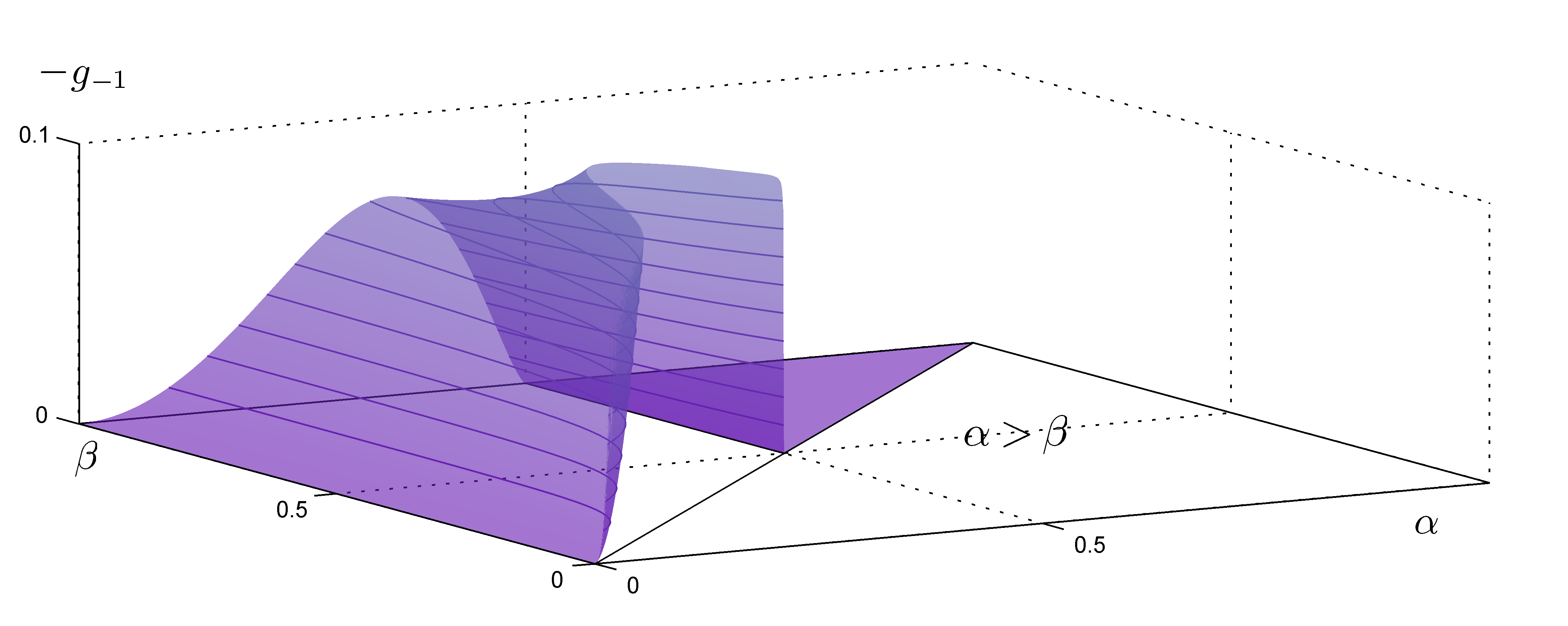}
\caption{Plot of the residue $g_{-1}$ at $z = z_0(\alpha)$. This residue vanishes moving into the low density phase. In the region $\alpha > \beta$, a different singularity dominates. In the high density phase $\alpha > \beta$, $\beta < \frac{1}{2}$, the pole at $z = z_0(\beta)$ is closer to the origin.}
\end{figure}

The vanishing of the residue at $\alpha = 1/2$, shown in Figure~\ref{fig:poleresidue}, indicates that $\mathcal{Q}(z;\alpha,\beta)$ is well behaved at $z_0(\alpha)$ at this point in the maximal current phase. When probing further we find beyond $\alpha = 1$, $\beta = 1$ that $z_0(\alpha)$ again becomes a pole, however not as the singularity closest to the origin.

Focusing on the $\alpha<1/2$ region, knowing the position and magnitude of the pole, along with the residue, we use \eref{eq:coeffscale1} to find the asymptotic scaling of the sum of squared weights
\begin{equation} \label{eq:squaredweightsLD}
\sum_{C\in\mathcal{L}_N}\mathcal{W}({\cal C})^2 \sim -\frac{g_{-1}}{z_0(\alpha)} \left(\frac{\alpha^2+(1-\alpha)^2}{\alpha^2(1-\alpha)^2}\right)^N.
\end{equation}
To normalise these squared weights into squared \emph{probabilities}, we divide through \eref{eq:squaredweightsLD} by the normalisation $Z_N(\alpha, \beta)$ \eref{eq:ZN}, squared. Within the low density phase we know the asymptotic form of this to be \cite{Blythe2007}
\begin{equation}
\label{eq:ZLD}
Z_N(\alpha, \beta) \sim \frac{\beta(1-2\alpha)}{(\beta-\alpha)(1-\alpha)}\left(\frac{1}{\alpha(1-\alpha)}\right)^N
\end{equation}
from which we obtain the sum of squared probabilities,
\begin{equation}
\label{eq:sumsquareprobsLD}
\sum_{C\in\mathcal{L}_N} P({\cal C})^2 \sim \frac{g_{-1}(1-\alpha)^2(\beta-\alpha)^2}{\beta^2(1-2\alpha)^2}\left(\alpha^2+(1-\alpha)^2\right)^N.
\end{equation}
The $\lambda = 2$ R\'{e}nyi entropy $H_2$ follows, which to leading order is
\begin{equation}
\label{eq:H2LD}
H_2 = -\log{\sum_{C\in\mathcal{L}_N} P({\cal C})^2} \sim -N\log{\left(\alpha^2+(1-\alpha)^2\right)} + \mathcal{O}(1)
\end{equation}
and an effective number of configurations with $\e^{H_2}$
\begin{equation}
\label{eH2}
\e^{H_2} \sim \frac{\beta^2(1-2\alpha)^2}{g_{-1}(1-\alpha)^2(\beta-\alpha)^2}\left(\frac{1}{\alpha^2+(1-\alpha)^2}\right)^N.
\end{equation}

\subsection{High density phase $\beta < \alpha$, $\beta < \frac{1}{2}$}
By the symmetry of the generating function and of the dynamics of particles and holes in the TASEP, the corresponding results in the high density phase are an $(\alpha,\beta) \rightarrow (\beta,\alpha)$ mirror of those found in the low density phase.

\subsection{Maximal current phase $\alpha > \frac{1}{2}$, $\beta > \frac{1}{2}$}
We find in this phase the dominant singularity to be a branch point, at $z_1 = 1/8$. A series expansion of $\mathcal{Q}(z;\alpha,\beta)$ about this branch point shows the emergence of an imaginary contribution of order $(z-z_1)^\frac{3}{2}$:
\begin{equation}
\mathcal{Q}\left(z;\alpha,\beta\right) = ih_{\frac{3}{2}}(\alpha,\beta)\left(z-\frac{1}{8}\right)^\frac{3}{2} + \sum_{j\geq 0}h_j\left(z-\frac{1}{8}\right)^j.
\end{equation}
This is a signature of an algebraic singularity of order $(z-z^*)^\frac{3}{2}$. We find $h_{\frac{3}{2}}$, that we quote in \ref{S:Residues} \eref{h32}. Using \eref{eq:coeffscale2}, we find the asymptotic scaling of the sum of squared weights
\begin{equation} \label{eq:squaredweightsMC}
\sum_{C\in\mathcal{L}_N}\mathcal{W}({\cal C})^2 \sim \frac{h_{\frac{3}{2}}\left(\frac{1}{8}\right)^\frac{3}{2}}{\Gamma(-\frac{3}{2})} \frac{8^N}{N^{\frac{5}{2}}} \;.
\end{equation}
\begin{figure}[t]
\includegraphics[width=1\textwidth]{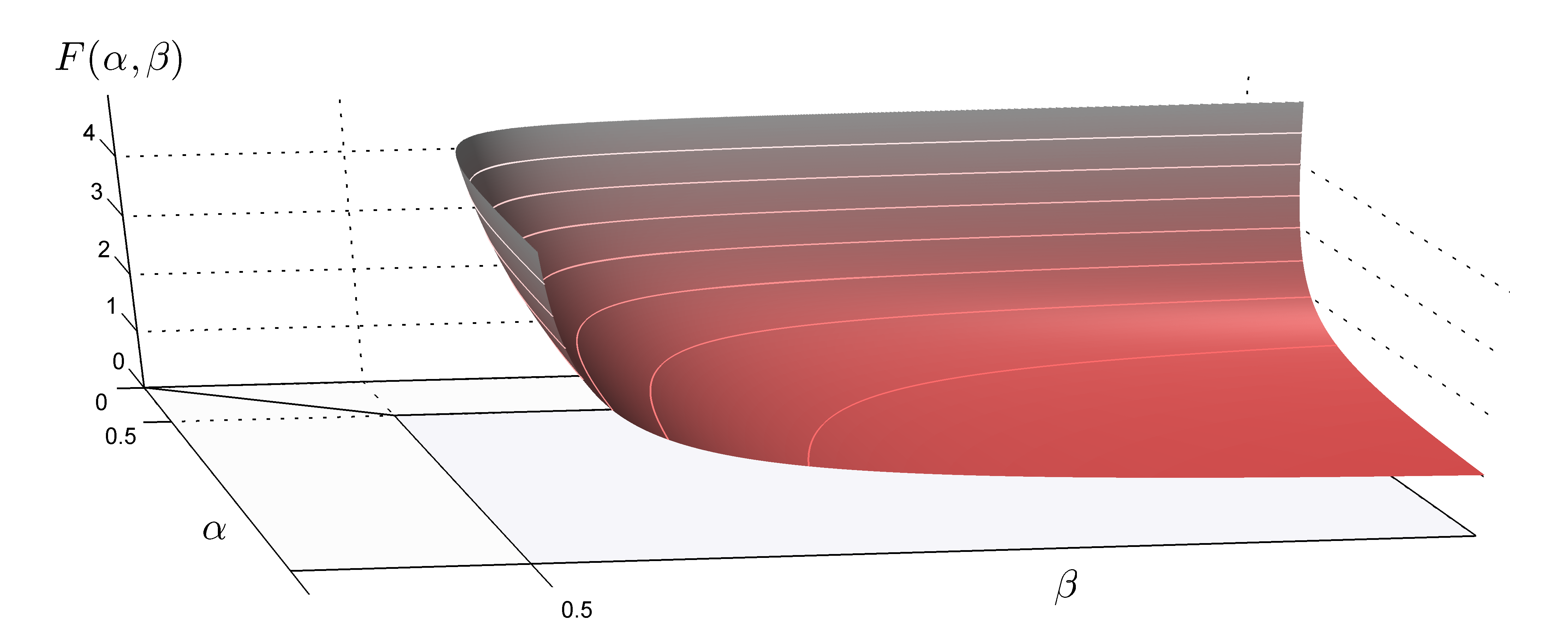}
\caption{Surface plot of $F(\alpha,\beta).$ Going deeper into the maximal current phase, the effective number of participating configurations $\e^{H_2}$ decreases, towards a constant. $F$ is divergent on the phase boundaries, however the effective number is well behaved for all $\alpha, \beta > \frac{1}{2}$.}
\label{fig:effno}
\end{figure}
We normalise this to obtain the sum of squared probabilities using the appropriate asymptotic expression for the normalisation $Z_N(\alpha, \beta)$ \eref{eq:ZN} in this phase \cite{Blythe2007}
\begin{equation}
Z_N(\alpha, \beta) \sim \frac{4\alpha\beta(\alpha+\beta-1)}{\sqrt{\pi}(2\alpha-1)^2(2\beta-1)^2}\frac{4^N}{N^{\frac{3}{2}}}
\end{equation}
to obtain a final expression
\begin{eqnarray}
\label{eq:sumsquareprobsMC}
\sum_{C\in\mathcal{L}_N} P({\cal C})^2 &\sim \frac{1}{F(\alpha,\beta)}\sqrt{\frac{\pi}{2}}\frac{\sqrt{N}}{2^N}
\end{eqnarray}
where the prefactor $F$ (which we quote in \eref{eq:effno}) has no dependence on system size. Thus the large $N$ scaling of $H_2$ is
\begin{equation}
\label{eq:H2MC}
H_2 = -\log{\sum_C P({\cal C})^2} \sim N\log{2} - \frac{1}{2}\log N + \mathcal{O}(1) \;.
\end{equation}
For large system sizes, the leading contributions to the R\'{e}nyi entropy becomes independent  of $\alpha$, $\beta$. For the effective number of configurations $\e^{H_2}$, however, $\alpha$ and $\beta$ arise in the multiplicative factor $F(\alpha,\beta)$ 
\begin{equation}
\label{eq:eH2MC}
\e^{H_2} \sim F(\alpha,\beta)\sqrt{\frac{2}{\pi}}\frac{2^N}{\sqrt{N}} \;.
\end{equation}
To interpret this scaling with system size, recall that the maximal current phase has bulk density $\rho = 1/2$. Consider now the asymptotic form of the binomial coefficient
\begin{equation}
{{N}\choose{\frac{N}{2}}}\sim \sqrt{\frac{2}{\pi}}\frac{2^N}{\sqrt{N}}
\end{equation}
and note the same scaling with $N$ as the effective number \eref{eq:eH2MC}. 

Illustrated in Figure~\ref{fig:effno}, the prefactor $F$ is a decreasing function of $\alpha$ and $\beta$. For some special cases we obtain neat results for this effective number:
\begin{eqnarray}
\e^{H_2}(\alpha,\alpha) \sim \frac{\sqrt{6}\left(\sqrt{3}+2 \alpha -1\right)^2}{2(2 \alpha -1)
   \left(\sqrt{3}+2 (2 \alpha -1)\right)}  \left[\sqrt{\frac{2}{\pi}}\frac{2^N}{\sqrt{N}}\right] \\
\e^{H_2}(1,1) \sim \sqrt{6} \left[\sqrt{\frac{2}{\pi}}\frac{2^N}{\sqrt{N}}\right] \\
\e^{H_2}(\infty,\infty) \sim \frac{1}{4}\sqrt{6} \left[\sqrt{\frac{2}{\pi}}\frac{2^N}{\sqrt{N}}\right].
\end{eqnarray}

\subsection{$H_2$ phase diagram}
\begin{figure}[t]
\includegraphics[trim=0mm 5mm 0mm 0mm]{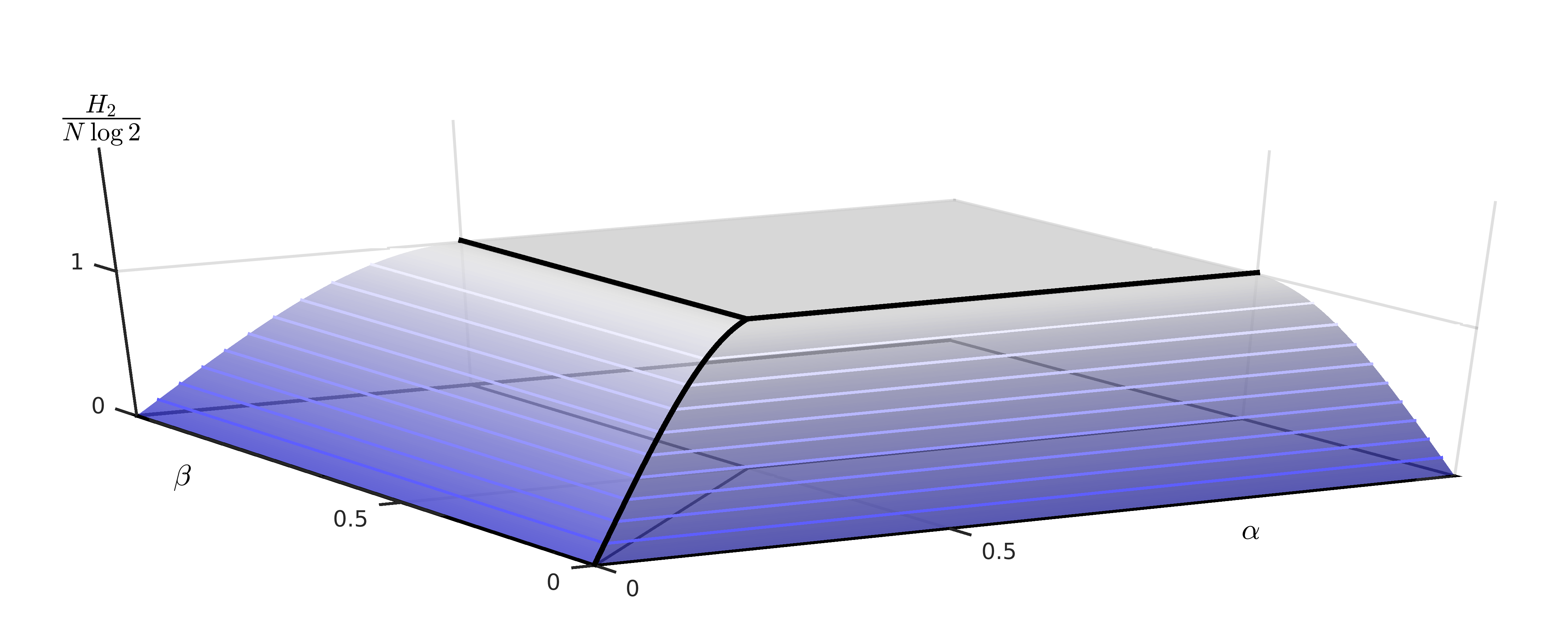}
\caption{Asymptotic scaling of the R\'{e}nyi entropy $H_2$, across the phase diagram.}
\label{fig:h2phasediag}
\end{figure}

These results are summarised with a surface plot of $H_2$ across the phase diagram in Figure~\ref{fig:h2phasediag}. We find a plateau in this R\'{e}nyi entropy in the maximal current phase, that arises from the branch point with no $\alpha$, $\beta$ dependence.

\subsection{Bounds on Shannon entropy}
The R\'{e}nyi entropy is a nonincreasing function of $\lambda$. With our results for $H_2$, and knowing that $H_0 = N\log{2}$ across the whole phase diagram, we find bounds on the Shannon entropy in the three phases
\begin{eqnarray}
\mathrm{Low} \ \mathrm{density}\quad & N\log{2} &\geq S(\alpha,\beta) \geq -N\log{\left(\alpha^2 + (1-\alpha)^2\right)}   \\
\mathrm{High} \ \mathrm{density}\quad & N\log{2} &\geq S(\alpha,\beta) \geq -N\log{\left(\beta^2 + (1-\beta)^2\right)}   \\
\mathrm{Maximal} \ \mathrm{current} \quad & N\log{2} &\geq S(\alpha,\beta) \geq N\log{2} - \frac{1}{2}\log{N}.
\end{eqnarray}

\section{Discussion and conclusion}
\label{S:disc}
In this work we have derived an exact expression \eref{eq:Qzab} for the generating function of the sum of squared weights for the open TASEP. In the particular case  $\alpha = \beta =1$ this simplifies and allows 
a finite sum expression for the sum of squared weights \eref{eq:a1b1}.

From these expressions we derive the large $N$ behaviour of the $\lambda = 2$ R\'{e}nyi entropy \eref{eq:H2LD}, \eref{eq:H2MC}. As we shall discuss, the leading order terms in these entropies are what one would obtain from a Bernoulli measure - the system at the same particle density, with correlations absent. While one may anticipate this as the leading order term, the corrections to this order are reflective of correlations in the NESS, which take different forms in the different phases. These in turn give the effective number of participating configurations \eref{eH2} \eref{eq:eH2MC}. 

In the high and low density phases, it is known that density-density correlations decay exponentially with distance. In turn, we find the correction to the Bernoulli measure expression to be $\mathcal{O}(1)$. In the maximal current phase, however, there is a long range power law decay: for sites $i$ and $j$ with occupations $\tau$,  $\langle (\tau_i-\frac{1}{2})(\tau_j -\frac{1}{2}) \rangle \sim |i-j|^{-\frac{1}{2}}$\cite{derrida1993b}. We in turn find an $\mathcal{O}(\log N)$ correction in this phase. These corrections represent non-additive contributions to the R\'enyi entropy. 

It would be interesting to establish how the corrections to the R\'enyi entropy are intrinsically related to the nature of correlations - specifically, whether one can infer the correction to the R\'{e}nyi entropy of a system, from the correlations it exhibits.

\subsection{Asymptotic form of the R\'enyi entropy}

We saw that along the factorisation line $\alpha +\beta =1$ we can write down a simple expression for all R\'{e}nyi entropies \eref{eq:MFrenyi}. This is simply the result of a Bernoulli measure for the stationary state. We notice that in the case $\lambda =2$ the same expression  gives the leading order term in the exact expressions \eref{eq:H2LD}, \eref{eq:H2MC}, when we take $\rho$ to be the density within the bulk of the system, $\rho = \alpha, 1-\beta, 1/2$ in the high density, low density and maximal current phases respectively. 

We anticipate that in the low density and high density phases, the leading scaling with system size $N$ for all R\'{e}nyi entropies is given by a Bernoulli measure \eref{eq:MFrenyi}. We thus conjecture the following:
\begin{eqnarray}
\mathrm{Low} \ \mathrm{density}\quad & H_\lambda &= \frac{N}{1-\lambda}\log{\left(\alpha^\lambda + (1-\alpha)^\lambda\right)} + \mathcal{O}(1)  \label{ldconj} \\ 
& S & = -N\left(\alpha \log{\alpha} + (1-\alpha)\log{(1-\alpha)}\right) + \mathcal{O}(1)
\\
\mathrm{High} \ \mathrm{density}\quad & H_\lambda &= \frac{N}{1-\lambda}\log{\left(\beta^\lambda + (1-\beta)^\lambda\right)} + \mathcal{O}(1)  \label{hdconj}  \\
& S &= -N\left(\beta \log{\beta} + (1-\beta)\log{(1-\beta)}\right) + \mathcal{O}(1) \; . 
\end{eqnarray}

Within the maximal current phase we conjecture that for \emph{all} $\lambda \geq 1$, the leading behaviours are 
\begin{eqnarray}
H_\lambda &\sim N\log{2} - \frac{1}{2}\log{N} + \mathcal{O}(1) \\
\e^{H_\lambda} &\propto \frac{2^N}{\sqrt{N}} \label{mcconj2}
\end{eqnarray}
so that the leading correction is logarithmic in system size with the prefactor $1/2$ arising from the square root.
To understand this conjecture we note that the behaviour
of the effective numbers \eref{mcconj2} would be the asymptotic scaling of the binomial coefficient ${{N}\choose{N/2}}$. This binomial coefficient gives the number of half-filled configurations, which are the configurations expected to dominate the maximal current phase.

As a check of the conjecture for the high and low density phases (\ref{ldconj},\ref{hdconj}) we consider the $\lambda \to \infty$ entropy, $H_\infty$. Taking this limit for the case of a Bernoulli measure \eref{eq:MFrenyi} we find
\begin{equation}
\label{eq:MFinfrenyi}
H_\infty=
\cases{ -N\log{\left(1-\rho\right)} & \mbox{if} $\rho < \frac{1}{2}$\\
	-N \log{\rho} & \mbox{if} $\rho > \frac{1}{2}$.}
\end{equation}
Generally, assuming no degeneracy in the maximum probability $\max{\{P({\cal C})\}}$ within the distribution, $H_\infty$ is equal to $-\log\left(\max{\{P({\cal C})\}}\right)$. In the low density phase of the TASEP, for large system sizes, the most probable of the $2^N$ available configurations is an entirely empty system, that has weight $\mathcal{W} = \bra{W}E\cdots E\ket{V} = \left(1/\alpha\right)^N$. By normalising the weight of this empty configuration with \eref{eq:ZLD}, we find $H_\infty$ in this phase to be 
\begin{equation}
H_\infty \sim -N\log(1-\alpha) - \log{\frac{(1-\alpha)(\beta-\alpha)}{\beta(1-2\alpha)}}
\end{equation}
thus the leading order term is the Bernoulli measure result \eref{eq:MFinfrenyi}. The high density result is obtained by similar means, where the most probable configuration is one with every site occupied.
Thus the Bernoulli measure result correctly gives the leading order term for the cases $\lambda =  2, \infty$ (and trivially $\lambda = 0$).

\subsection{Absence of two transition lines: Lee-Yang zeros}

\begin{figure}[t]
\centering
\includegraphics[width=\textwidth, trim=5mm 0mm 5mm 0mm]{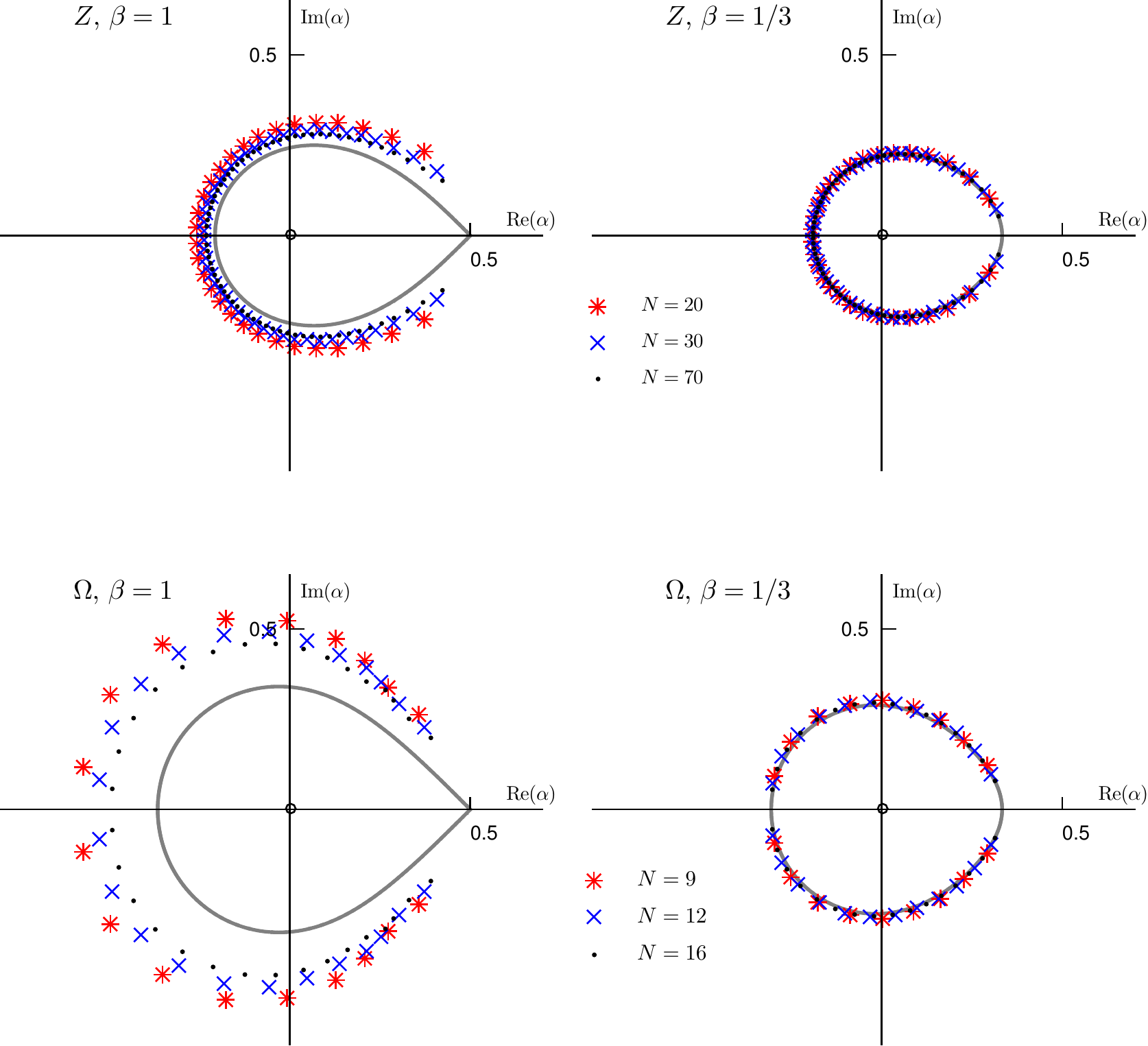}
\caption{Lee-Yang zeros of the sum of weights $Z$, and sum of squared weights $\Omega$ for differing system sizes $N$. $\Omega$ must be computed from a series expansion of the generating function \eref{eq:Qzab}, while $Z$ is found directly using \eref{ZNexact}.}
\label{fig:leeyang}
\end{figure}

Let us write the R\'{e}nyi entropy for the case  $\lambda = 2$ in the form 
\begin{equation}
\label{eq:renyieq2}
H_2 = - \log{\Omega} + 2 \log{Z}
\end{equation}
where  $Z$ is usual the sum of weights, and we have introduced $\Omega$ to denote the sum of squared weights. Now in the special case of thermal equilibrium \eref{eq:renyieqZ} where we have Boltzmann weights with temperature $T$, we find $\Omega = Z\left(T/2\right)$. Therefore, two nonanalyticities emerge in $H_2$ arising from a phase transition at critical temperature $T^* $. The first is at $T = T^*$, coming from $Z$ and a second at $T = 2T^*$ coming from the special form of $\Omega$.

For the TASEP, however, we find instead that $\Omega$ exhibits nonanalyticities at the same parameter values as  $Z$ in \eref{eq:renyieq2}, and no secondary transition lines arise. This result is consistent with the TASEP being a far-from-equilibrium system with no reasonable notion of a temperature, and supports the idea of the equilibrium distribution being the special case.

It is, in principle, possible that there are additional nonanalyticities buried in \eref{eq:Qzab} that might correspond to a secondary phase transition. To exclude this possibility, we consider the Lee-Yang zeros of $Z$ and $\Omega$ which should reveal all such singularities. Indeed, this has been demonstrated previously for the TASEP normalisation \cite{Blythe2002,Blythe2003} (see also Figure~\ref{fig:leeyang}). Specifically, the normalisation \eref{ZNexact} takes the form of  a  polynomial in the transition rates $\alpha$ and $\beta$ (rather than Boltzmann factors $\e^{-1/T}$ as in the equilibrium case). If, say, we take fixed $\beta$, we plot the zeros of this polynomial in the complex $\alpha$ plane, to find they lie on a closed curve. The value of $\alpha$ at which the zeros pinch the real axis as $N\to \infty$ represents a critical value \cite{Blythe2002}. 

We have repeated this procedure for the sum of squared weights. As shown in Figure~\ref{fig:leeyang}, we find that in the case $\beta = 1$, both the roots of the normalisation and the sum of squared weights pinch the real $\alpha$ axis at $\alpha = 1/2$ (i.e.~the transition point between the low-density and maximal current phases). For $\beta = 1/3$, the roots approach $\alpha = 1/3$ (the high density-low density transition line). This serves as a verification that there are no further transition lines in $\Omega$. 

Using the method set out in \cite{Blythe2003} and the asymptotic expressions for the sum of squared weights, \eref{eq:squaredweightsLD} and \eref{eq:squaredweightsMC}, we would predict the zeros to approach the curve
\begin{equation}
\left| \frac{\alpha^2 + (1-\alpha)^2}{\alpha^2(1-\alpha)^2} \right| = \left\{
\begin{array}{ll}
\frac{\beta^2 + (1-\beta)^2}{\beta^2(1-\beta)^2} & \mbox{for $\beta<\frac{1}{2}$} \\
8 & \mbox{for $\beta>\frac{1}{2}$}
\end{array}\right.
\end{equation}
in the complex-$\alpha$ plane. These lines are plotted in Figure~\ref{fig:leeyang}, showing evidence that the zeros of the finite-degree polynomials are approaching the asymptotic curve. This provides an independent check that the asymptotic analysis of Section~\ref{S:asymp} is correct, and therefore is strong evidence that the secondary transitions that occur generically in equilibrium statistical mechanical systems are absent in the TASEP, and A natural question is whether there is a deep underlying reason for this to be the case. 

We conclude with the remark that for one-dimensional quantum systems
the scaling of the R\'enyi entropy (defined in terms of a density
matrix) indicates whether a matrix product state will be an efficient
ansatz for the simulation of the system \cite{Schuch2008}. It would be
of interest to determine whether similar considerations hold for
classical nonequilibrium states.  Indeed further insights may come
from the study of the R\'{e}nyi entropy in the context of a wider
range of nonequilibrium states.

\ack{AJW acknowledges studentship funding from EPSRC under grant number EP/L015110/1 and MRE acknowledges funding under grant number EP/J007404/1. }
\\

\bibliographystyle{iopart-num}
\bibliography{library}

\appendix

\section{Extraction of coefficients for obtaining $\mathcal{R}(t;a,a,b,b)$}
\label{S:xextraction}
In this section we demonstrate the calculation of extracting coefficients to obtain, from the recursion relation \eref{eq:mainrecurrence}, a closed form expression for the generating function $\mathcal{R}(t;a,a,b,b)$, that for the number of walks beginning and terminating at points on the diagonal. We return to expression \eref{eq:Rxy1}, which we restate for convenience
\begin{eqnarray} 
\fl xy\mathcal{R}(x,y) +\bar{y}\mathcal{R}(x,\bar{x}\bar{y}) - \bar{x}\mathcal{R}(\bar{x}\bar{y},y) = \frac{1}{\sqrt{\Delta(x)}}\left[\frac{1}{1-\bar{y}Y_-(x)}+\frac{1}{1-y\bar{Y}_+(x)}-1\right] \times
\\\fl \left[\frac{xy}{(1-bx)(1-by)} +\frac{\bar{y}}{(1-bx)(1-b\bar{x}{\bar{y}})} - \frac{\bar{x}}{(1-b\bar{x}\bar{y})(1-by)}+t\mathcal{R}(0,0)-2t(1+x)\mathcal{R}(x,0)\right] \nonumber.
\end{eqnarray}

\subsection{$y^0$ coefficient extraction}

Extraction of the $y^0$ component is this time more involved. If we take a cross term of \eref{eq:Rxy1} as an example, we obtain the $y^0$ coefficient with an explicit series expansion
\begin{eqnarray}
&\left[\frac{1}{1-\bar{y}Y_-(x)}+\frac{1}{1-y\bar{Y}_+(x)}-1\right]\left[\frac{xy}{(1-bx)(1-by)}\right] \nonumber \\
& = \frac{x}{1-bx}\left[\sum_{n\geq 0}(\bar{y}Y_-)^n+\sum_{m\geq 0}(y\bar{Y}_+)^m-1\right]\left[y\sum_{p\geq 0}(by)^p\right] \nonumber \\
& = \frac{x}{1-bx}\sum_{n\geq 1}\sum_{p\geq 0} y^{1+p-n}Y_-^nb^p + \mathcal{O}(y) \nonumber \\
& = \frac{x}{1-bx}\sum_{n\geq 1} Y_-^{n}b^{n-1} + \mathcal{O}(y)+ \mathcal{O}(\bar{y}) \nonumber \\
& = \frac{xY_-(x)}{(1-bx)(1-bY_-(x))} + \mathcal{O}(y) + \mathcal{O}(\bar{y}). 
\end{eqnarray}
Applying this same method throughout, the $y^0$ component of \eref{eq:Rxy1} is
\begin{eqnarray}
\fl -\bar{x}R_D(\bar{x}) = & \frac{1}{\sqrt{\Delta(x)}}\Bigg[t\mathcal{R}(0)-2t(1+x)\mathcal{R}(x)+\frac{2xY_-(x)}{(1-bx)(1-bY_-(x))} \\
&-\frac{2b\bar{x}Y_-(x)}{(1-b^2\bar{x})(1-bY_-(x))}-\frac{\bar{x}}{1-b^2\bar{x}}\Bigg] \nonumber
\end{eqnarray}
where we have further condensed the notation and introduced $R_D(\bar{x}) \equiv \mathcal{R}(\bar{x},\bar{x})$, $\mathcal{R}(x) \equiv \mathcal{R}(x,0)$ and $\mathcal{R}(0) \equiv \mathcal{R}(0,0)$. Considering the explicit form of $Y_-(x)$ \eref{eq:Y0Y1}, one can rearrange the term
\begin{equation}
\frac{Y_-(x)}{1-bY_-(x)} = -\frac{1}{2}\frac{t(2b+1+2bx+x^2)-x+x\sqrt{\Delta(x)}}{(1+b)tx^2+(t+b(bt-1))x+b(1+b)t}
\end{equation}
whereby the quadratic in the denominator factorises in a way similar to the kernel
\begin{eqnarray}
\fl (1+b)tx^2+(t+b(bt-1))x+b(1+b)t = (1+b)t(x-bY_-(b,t))(x-bY_+(b,t)).
\end{eqnarray}
Using this, the full equation may be rearranged into the form
\begin{eqnarray}
\label{eq:beforex0}
\fl \sqrt{\Delta_+(\bar{x})}\left[-(1-bY_-(b)\bar{x})R_D(\bar{x})+\frac{x}{(1+b)t(x-bY_+(b))}\left(\frac{x}{1-bx}-\frac{b\bar{x}}{1-b^2\bar{x}}\right)\right] \\ 
\fl =\frac{1}{\sqrt{\Delta_0\Delta_+(x)}}\Bigg[(x-bY_-(b))\left(t\mathcal{R}(0)-2t(1+x)\mathcal{R}(x)\right) \nonumber \\ 
-\frac{t(2b+1+2bx+x^2)-x}{(1+b)t(x-bY_+(b))}\Bigg(\frac{x}{1-bx}-\frac{b\bar{x}}{1-b^2\bar{x}}\Bigg) - \frac{1-bY_-(b)\bar{x}}{1-b^2\bar{x}}\Bigg] \nonumber
\end{eqnarray}
where we have also multiplied through by a factor $(x-bY_-(b))$. At this point we note the impact of introducing the additional factors of $b$ into this calculation. Comparing \eref{eq:beforex0} to the $\beta = 1$ case \eref{eq:Pxy4}, we encounter a significantly more involved expression for this case, that we aim to extract the positive powers of $x$ from. 

\subsection{$x^+$ coefficient extraction}
We now find the $x^+$ coefficient of \eref{eq:beforex0}. To do this first we split this equation into six terms (suppressing the explicit dependence of $(b,t)$ in $Y_+(b,t)$, $Y_-(b,t)$)
\begin{eqnarray}
\fl \eref{eq:beforex0}: \quad T_1 + T_2 = T_3 + T_4 + T_5 + T_6 \\ 
\fl T_1 = -\sqrt{\Delta_+(\bar{x})}(1-bY_-\bar{x})R_D(\bar{x})  \\ 
\fl T_2 = \frac{\sqrt{\Delta_+(\bar{x})}}{(1+b)t}\frac{x}{x-bY_+}\left(\frac{x}{1-bx}-\frac{b\bar{x}}{1-b^2\bar{x}}\right)  \\
\fl T_3 = \frac{1}{\sqrt{\Delta_0\Delta_+(x)}}(x-bY_-)\left(t\mathcal{R}(0)-2t(1+x)\mathcal{R}(x)\right) \\
\fl T_4 = -\frac{1}{\sqrt{\Delta_0\Delta_+(x)}}\left(\frac{t(2b+1+2bx+x^2)-x}{(1+b)t(x-bY_+)}\left[\frac{x}{1-bx}\right]  \right)  \\
\fl T_5 = \frac{1}{\sqrt{\Delta_0\Delta_+(x)}}\left(\frac{t(2b+1+2bx+x^2)-x}{(1+b)t(x-bY_+)}\left[\frac{b\bar{x}}{1-b^2\bar{x}}\right]\right)  \\
\fl T_6 = - \frac{1}{\sqrt{\Delta_0\Delta_+(x)}}\frac{1-bY_-\bar{x}}{1-b^2\bar{x}} .
\end{eqnarray}
We now extract coefficients term by term. In this we use a number of identities involving the discriminant, found by considering formal power series, that we quote here: 
\begin{eqnarray}
\fl \left\{x^+\right\} \frac{\bar{x}^2}{\sqrt{\Delta_+(x)}(1-c\bar{x})} \\
= \frac{1}{1-c\bar{x}}\left(\frac{\bar{x}^2}{\sqrt{\Delta_+(x)}} - \frac{\bar{c}^2}{\sqrt{\Delta_+(c)}} -(\bar{x}^2-\bar{c}^2) - \frac{1}{2}(X_-+X_+)(\bar{x}-\bar{c})\right) \nonumber \\
\fl \left\{x^+\right\} \frac{\bar{x}}{\sqrt{\Delta_+(x)}(1-c\bar{x})} = \frac{1}{1-c\bar{x}}\left(\frac{\bar{x}}{\sqrt{\Delta_+(x)}} - \frac{\bar{c}}{\sqrt{\Delta_+(c)}} -(\bar{x}-\bar{c})\right)  \\
\fl \left\{x^{+}\right\}\frac{1}{\sqrt{\Delta_+(x)}(1-c\bar{x})} = \frac{1}{1-c\bar{x}}\left(\frac{1}{\sqrt{\Delta_+(x)}}-\frac{1}{\sqrt{\Delta_+(c)}}\right)  \\
\fl \left\{x^+\right\} \frac{x}{\sqrt{\Delta_+(x)}(1-c\bar{x})} =  \frac{1}{1-c\bar{x}}\left(\frac{x}{\sqrt{\Delta_+(x)}}-\frac{c}{\sqrt{\Delta_+(c)}}\right)  \\
\fl \left\{x^+\right\} \frac{x^2}{\sqrt{\Delta_+(x)}(1-c\bar{x})} = \frac{1}{1-c\bar{x}}\left(\frac{x^2}{\sqrt{\Delta_+(x)}}-\frac{c^2}{\sqrt{\Delta_+(c)}}\right)  \\
\fl \left\{x^+\right\} \frac{x\sqrt{\Delta_+(\bar{x})}}{1-cx} = \frac{x\sqrt{\Delta_+(c)}}{1-cx}.
\end{eqnarray}
Again, $\left\{x^+\right\}$ denotes `the positive powers in $x$ within'. Here, $c$ may be any constant independent of $x$. Applying these, we now find the $\{x^+\}$ component of each term.

\begin{equation} 
\label{eq:term1}
\fl \left\{x_+\right\}T_1 = 0
\end{equation}

\begin{equation} 
\label{eq:term2}
\fl \left\{x^+\right\}T_2 = \frac{x\sqrt{\Delta_+(b)}}{t(1+b)(1-b^2Y_+)(1-bx)}
\end{equation}

\begin{eqnarray} 
\label{eq:term3}
\fl \left\{x^+\right\}T_3 & = \left(\frac{x-bY_-}{\sqrt{\Delta_0\Delta_+(x)}}-\frac{bY_-}{\sqrt{\Delta_0}}\right)t\mathcal{R}(0) - \left(\frac{2t(1+x)(x-bY_-)}{\sqrt{\Delta_0\Delta_+(x)}}\right)\mathcal{R}(x)
\end{eqnarray}

\begin{eqnarray} 
\label{eq:term4}
\fl\left\{x^+\right\}T_4 = -\frac{1}{\sqrt{\Delta_0}(1+b)(1-b^2Y_+)t}\Bigg\{\left[(2b+1)t\left(\frac{1}{(1-bx)\sqrt{\Delta_+(x)}}-1\right)\right] \\
\fl +\left[\frac{(2bt-1)x}{(1-bx)\sqrt{\Delta_+(x)}}\right]+\left[\frac{tx^2}{(1-bx)\sqrt{\Delta_+(x)}}\right]+\left[\frac{(2b+1)t}{1-bY_+\bar{x}}\left(\frac{1}{\sqrt{\Delta_+(x)}}-\frac{1}{\sqrt{\Delta_+(bY_+)}}\right)\right]\nonumber \\
\fl + \left[\frac{2bt-1}{1-bY_+\bar{x}}\left(\frac{x}{\sqrt{\Delta_+(x)}}-\frac{bY_+}{\sqrt{\Delta_+(bY_+)}}\right)\right]\nonumber + \left[\frac{t}{1-bY_+\bar{x}}\left(\frac{x^2}{\sqrt{\Delta_+(x)}}-\frac{b^2Y_+^2}{\sqrt{\Delta_+(bY_+)}}\right)\right] \nonumber\\
\fl -\left[(2b+1)t\left(\frac{1}{\sqrt{\Delta_+(x)}}-1\right)\right] - \left[\frac{(2bt-1)x}{\sqrt{\Delta_+(x)}}\right]-\left[\frac{tx^2}{\sqrt{\Delta_+(x)}}\right]\Bigg\} \nonumber
\end{eqnarray}

\begin{eqnarray} 
\label{eq:term5}
\fl \left\{x^+\right\}T_5 =  \frac{1}{\sqrt{\Delta_0}(1+b)(Y_+-b)t}\Bigg\{bY_+\Bigg[\frac{(2b+1)t}{1-bY_+\bar{x}}\Bigg(\frac{\bar{x}^2}{\sqrt{\Delta_+(x)}}-\frac{Y_-^2}{\sqrt{\Delta_+(bY_+)}} \\
\fl -(\bar{x}^2-Y_-^2)-\frac{1}{2}(\bar{x}-Y_-)(X_-+X_+)\Bigg) + \frac{2bt-1}{1-bY_+\bar{x}}\Bigg(\frac{\bar{x}}{\sqrt{\Delta_+(x)}}-\frac{Y_-}{\sqrt{\Delta_+(bY_+)}} \nonumber \\
\fl -(\bar{x}-Y_-)\Bigg)+\frac{t}{1-bY_+\bar{x}}\Bigg(\frac{1}{\sqrt{\Delta_+(x)}}-\frac{1}{\sqrt{\Delta_+(bY_+)}}\Bigg)\Bigg] -b^2\Bigg[\frac{(2b+1)t}{1-b^2\bar{x}}\Bigg(\frac{\bar{x}^2}{\sqrt{\Delta_+(x)}} \nonumber \\
\fl -\frac{\bar{b}^4}{\sqrt{\Delta_+(b^2)}}-(\bar{x}^2-\bar{b}^4)-\frac{1}{2}(\bar{x}-\bar{b}^2)(X_-+X_+)\Bigg) + \frac{2bt-1}{1-b^2\bar{x}}\Bigg(\frac{\bar{x}}{\sqrt{\Delta_+(x)}}-\frac{\bar{b}^2}{\sqrt{\Delta_+(b^2)}} \nonumber \\
\fl -(\bar{x}-\bar{b}^2)\Bigg)+\frac{t}{1-b^2\bar{x}}\left(\frac{1}{\sqrt{\Delta_+(x)}}-\frac{1}{\sqrt{\Delta_+(b^2)}}\right)\Bigg]\Bigg\} \nonumber 
\end{eqnarray}

\begin{equation} 
\label{eq:term6}
\fl \left\{x^+\right\}T_6 = - \frac{1}{\sqrt{\Delta_0}(1-b^2\bar{x})}\left(\frac{1-bY_-\bar{x}}{\sqrt{\Delta_+(x)}}-\frac{1-\bar{b}Y_-}{\sqrt{\Delta_+(b^2)}}+bY_-(\bar{x}-\bar{b}^2)\right).
\end{equation}

With \eref{eq:term1} - \eref{eq:term6}, an explicit expression for $\mathcal{R}(x) = \mathcal{R}(t;x,0,b,b)$ is found with a rearrangement of terms. Using the recursion relation in equation \eref{eq:simplerecurrence}, and putting in an explicit form for $\mathcal{R}(t;a,0,b,b)$, we acquire a preliminary expression for $\mathcal{R}(t;a,a,b,b)$
\begin{eqnarray} 
\label{eq:Raatbb1}
\fl \frac{(a-b Y_-) a^2 K_{aa}}{\sqrt{\Delta_0 \Delta_+(a)}}R(t;a,a,b,b) = \frac{a^2 (a-b Y_-)}{(1-a b)^2 \sqrt{\Delta_0 \Delta_+(a)}}+\frac{b Y_- t R(t;0,0,b,b)}{\sqrt{\Delta_0}} \\
\fl +\frac{1}{\sqrt{\Delta_0} (1+b) \left(1-b^2 Y_+\right) t}\Bigg\{\frac{a\sqrt{\Delta_0\Delta_+(b)}}{1-ab}+(2 b+1) t \left(\frac{1}{(1-ab) \sqrt{\Delta_+(a)}}-1\right) \nonumber \\ 
\fl +\frac{(2 b t-1) a}{(1-ab)\sqrt{\Delta_+(a)}}+\frac{t a^2}{(1-ab)\sqrt{\Delta_+(a)}}+\frac{(2 b+1) t}{1-\bar{a}bY_+}\left(\frac{1}{\sqrt{\Delta_+(a)}}-\frac{1}{\sqrt{\Delta_+(bY_+)}}\right) \nonumber \\
\fl +\frac{(2 b t-1)}{1-\bar{a}bY_+} \left(\frac{a}{\sqrt{\Delta_+(a)}}-\frac{b Y_+}{\sqrt{\Delta_+(bY_+)}}\right)+\frac{t }{1-\bar{a}b Y_+}\left(\frac{a^2}{\sqrt{\Delta_+(a)}}-\frac{(b Y_+)^2}{\sqrt{\Delta_+(bY_+)}}\right) \nonumber \\
\fl -(2 b+1) t \left(\frac{1}{\sqrt{\Delta_+(a)}}-1\right)-\frac{(2 b t-1) a}{\sqrt{\Delta_+(a)}}-\frac{t a^2}{\sqrt{\Delta_+(a)}}\Bigg\} \nonumber \\ 
\fl -\frac{b Y_+}{\sqrt{\Delta_0} (1+b) (Y_+-b) t}\Bigg\{\frac{(2b+1)t}{1-\bar{a}b Y_+}\Bigg[ \frac{\bar{a}^2}{\sqrt{\Delta_+(a)}}-\frac{Y_-^2}{\sqrt{\Delta_+(bY_+)}}-\left(\bar{a}^2-Y_-^2\right) \nonumber \\ 
\fl -\frac{1}{2}\left(\bar{a}-Y_-\right) (X_- + X_+)\Bigg]+\frac{(2 b t-1)}{1-\bar{a}b Y_+}\Bigg[ \frac{\bar{a}}{\sqrt{\Delta_+(a)}}-\frac{Y_-}{\sqrt{\Delta_+(bY_+)}}-\left(\bar{a}-Y_-\right)\Bigg] \nonumber \\ 
\fl +\frac{t}{1-\bar{a}b Y_+}\Bigg[\frac{1}{\sqrt{\Delta_+(a)}}-\frac{1}{\sqrt{\Delta_+(bY_+)}}\Bigg]\Bigg\} \nonumber \\
\fl + \frac{b^2}{\sqrt{\Delta_0} (1+b) (Y_+-b) t}\Bigg\{ \frac{(2 b+1) t}{1-\bar{a}b^2} \Bigg[\frac{\bar{a}^2}{\sqrt{\Delta_+(a)}}-\frac{\bar{b}^4}{\sqrt{\Delta_+(b^2)}}-\left(\bar{a}^2-\bar{b}^4\right) \nonumber \\ 
\fl -\frac{1}{2} \left(\bar{a}-\bar{b}^2\right)(X_- + X_+)\Bigg]+\frac{(2 b t-1)}{1-\bar{a}b^2}\left(\frac{\bar{a}}{\sqrt{\Delta_+(a)}}-\frac{\bar{b}^2}{\sqrt{\Delta_+(b^2)}}-\left(\bar{a}-\bar{b^2}\right)\right) \nonumber \\
\fl +\frac{t}{1-\bar{a}b^2}\left(\frac{1}{\sqrt{\Delta_+(a)}}-\frac{1}{\sqrt{\Delta_+(b^2)}}\right)\Bigg\}\nonumber \\
+\frac{1}{\sqrt{\Delta_0}\left(1-\bar{a}b^2\right)}\left(\frac{1-\bar{a}b Y_-}{\sqrt{\Delta_+(a)}}-\frac{1-\bar{b}Y_-}{\sqrt{\Delta_+(b^2)}}+b Y_- \left(\bar{a}-\bar{b}^2\right)\right). \nonumber
\end{eqnarray}
This is highly nested. By using the form of the factorised kernel \eref{eq:factorisedkernel}, however, with extensive algebra we acquire from \eref{eq:Raatbb1} a more concise expression
\begin{eqnarray} \label{eq:Raatbb2}
\fl a^2b^2K_{\bar{a}b}K_{aa}\mathcal{R}(t;a,a,b,b) =& \frac{abK_{aa}}{(1-ab)^2}-\frac{\sqrt{\Delta_0\Delta_+(a)\Delta_+(b)}(\bar{a}-Y_-)ab}{(1-ab)(b-Y_-)} \\
&- \frac{\sqrt{\Delta_+(a)\Delta(b)}\left[b^2-(1+2b-b^2)t-(1+b)^2tbY_+\right]}{\sqrt{\Delta_+(bY_+)}(1+b)bK_{bb}}\nonumber \\ 
&+\frac{\sqrt{\Delta_+(a)\Delta_+(b)}t(\bar{a}-Y_-)\left(b^2\sqrt{\Delta_0}-t\frac{t-(2b+1)\sqrt{\Delta_0}}{t-\sqrt{\Delta_0}}\right)}{bK_{bb}}\nonumber. 
\end{eqnarray}
We now use a symmetry property: reversing the start and end points of our six-path walk does not change the number of paths between them. This is seen in the TASEP weights, as an invariance on exchanging $\alpha$, $\beta$. We then know that $\mathcal{R}(t;a,a,b,b)$ and $\mathcal{R}(t;b,b,a,a)$ must be equivalent. Exploiting this, and simplifying a number of the nested square root expressions by the denesting formula in \eref{eq:denesting} we make \eref{eq:Raatbb2} manifestly symmetric in $a$, $b$
\begin{eqnarray} 
\label{eq:Raatbbfinal}
\fl \mathcal{R}(t;a,a,b,b) \\ 
\fl = \frac{1}{(1-ab)^2abK_{\bar{a}b}} - \frac{\sqrt{\Delta_+(a)\Delta_+(b)}(1+ab)t^2}{a^3b^3K_{aa}K_{bb}K_{\bar{a}b}} - \frac{\sqrt{\Delta_+(a)\Delta_+(b)}t}{(1-ab)a^2b^2K_{aa}K_{bb}\sqrt{X_-X_+}} \nonumber \\ 
\fl + \frac{\sqrt{\Delta_+(a)\Delta_+(b)}t}{a^3b^3K_{aa}K_{bb}K_{\bar{a}b}\sqrt{X_-X_+}}\left[(a+b)t\frac{\sqrt{X_-X_+}+1}{\sqrt{X_-X_+}-1}+\frac{ab}{2}\left(1+2t+\sqrt{1-4t-12t^2}\right)\right] \nonumber.
\end{eqnarray}

\section{Residues of the generating function $Q(z;\alpha,\beta)$}
\label{S:Residues}
In this section we quote results for the residues of the generating function $Q(z;\alpha,\beta)$, which we obtain with a series expansion about the relevant singularity. To begin, expanding about the pole $z_0(\alpha) = \alpha^2(1-\alpha)^2/(\alpha^2 + (1-\alpha)^2)$ strictly within the low density phase $\alpha < \beta$, $\alpha < 1/2$ as per \eref{eq:poleexpand} (elsewhere, either the generating function is analytic at $z_0(\alpha)$, or the pole is subdominant) we find a residue
\begin{eqnarray}
\label{eq:g1}
\fl g_{-1} = -\left(\frac{1-2 \alpha }{1-\alpha }\right)^{3/2}\frac{\beta^2}{\sqrt{\beta\left(\alpha^2+(1-\alpha)^2\right)-\alpha ^2}} \times \\ 
\fl \left[\frac{2 (1-\alpha ) \alpha ^2 (1-\alpha -\beta )+\left(2 \alpha ^2 \beta+\beta -\alpha  \beta -\alpha ^2-1\right) \left(1+\sqrt{1-4 (1-\alpha )^2 \alpha ^2}\right)}{4\sqrt{\alpha } (\alpha -\beta ) \left[(1-\alpha)\alpha +\left(\alpha ^2+(1-\alpha )^2\right) (1-\beta ) \beta \right] \left(\alpha ^2+(1-\alpha )^2\right)}\right]\times\nonumber \\
\fl \sqrt{2(1-\alpha)^2\alpha^2+(1-\beta)\left(\sqrt{1-4(1-\alpha)^2\alpha^2}-1\right)}\times \nonumber\\
\sqrt{2(1-\alpha)^2\alpha^2+(1-\alpha)\left(\sqrt{1-4(1-\alpha)^2\alpha^2}-1\right)}\nonumber.
\end{eqnarray}
About the branch point at $z_1 = 1/8$, within the maximal current phase $\alpha > 1/2$, $\beta > 1/2$ we find the residue
\begin{eqnarray}
\label{h32}
\fl h_{\frac{3}{2}} = \left[\frac{512\sqrt{2} \alpha ^2 \beta ^2 (1-\alpha -\beta )^2}{9 \left(1+2 \beta -2 \beta ^2\right) \left(1+2 \alpha -2 \alpha ^2\right) ((1-2 \alpha ) (1-2 \beta ))^{7/2}}\right]\times  \\
\left[\frac{1}{\sqrt{\left(112-64\sqrt{3}\right) \alpha\beta +\left(60 \sqrt{3}-104\right) (\alpha +\beta )-56 \sqrt{3}+97}}\right]\times \nonumber \\
\Bigg[\left(64 \sqrt{3}-96\right) \alpha ^2 \beta ^2 + \left(28\sqrt{3}-18\right)\alpha\beta  +\left(36-20 \sqrt{3}\right)\left(\alpha^2+\beta^2\right) \nonumber \\
+\left(48-40 \sqrt{3}\right) \left(\alpha ^2 \beta +\alpha  \beta ^2\right)+\left(11\sqrt{3}-21\right)(\alpha+\beta )+7\sqrt{3}-12\Bigg] \nonumber.
\end{eqnarray}
We quote the factor $F(\alpha,\beta)$ that we obtain when considering the effective number $\e^{H_2}$  \eref{eq:eH2MC} in the maximal current phase, plotted in Figure~\ref{fig:effno}
\begin{eqnarray}
\fl \frac{1}{F(\alpha,\beta)} = \frac{\sqrt{(1-2 \alpha ) (1-2 \beta )}}{3 \sqrt{2}\left(1+2\alpha-2\alpha^2\right) \left(1+2\beta-2\beta^2\right)}\times \\
\left[\frac{1}{\sqrt{\left(112-64\sqrt{3}\right) \alpha\beta +\left(60 \sqrt{3}-104\right) (\alpha +\beta )-56 \sqrt{3}+97}}\right] \times \nonumber \\
\Bigg[\left(64 \sqrt{3}-96\right) \alpha ^2 \beta ^2 + \left(28\sqrt{3}-18\right)\alpha\beta  +\left(36-20 \sqrt{3}\right)\left(\alpha^2+\beta^2\right) \nonumber \\
+\left(48-40 \sqrt{3}\right) \left(\alpha ^2 \beta +\alpha  \beta ^2\right)+\left(11\sqrt{3}-21\right)(\alpha+\beta )+7\sqrt{3}-12\Bigg] \nonumber.
\label{eq:effno}
\end{eqnarray}

\end{document}